\documentclass[draft]{agujournal2019}
\usepackage{url} 
\usepackage{lineno}
\usepackage[inline]{trackchanges} 

\usepackage{amsmath}
\usepackage{amssymb}
\usepackage[export]{}
\usepackage{lipsum} 

\usepackage{graphicx}

\usepackage{amsfonts}
\usepackage{fontenc}
\usepackage{cancel}
\usepackage{tensor}
\usepackage{rotating} 

\renewcommand\vec{\mathbf}

\newcommand*{\tensnd}[1]{\mathbb{#1}}

\newcommand*{\mean}[1]{\overline{{#1}}}
\DeclareMathSymbol{\mh}{\mathord}{operators}{`\-}

\usepackage{soul}
\draftfalse

\journalname{JGR: Space Physics}

\begin{document}

\title{Anomalous Electric Fields in Earth's Turbulent Magnetosheath: Insights From 3D Hybrid Simulations.}

\authors{J. A. Agudelo Rueda,\affil{1}, J. E. Stawarz,\affil{1}, L. Franci, \affil{1}, N. Yokoi, \affil{2}, C. Granier \affil{3,4,5}, and S.S. Cerri \affil{6}}

\affiliation{1}{School of Engineering, Physics and Mathematics, Northumbria University, Newcastle Upon Tyne, UK,}
\affiliation{2}{Earth and Planetary Science, University of Tokyo, Japan,}
\affiliation{3}{Department of Physics, Stanford University, 382 Via Pueblo Mall, Stanford, CA 94305, USA,}
\affiliation{4}{Kavli Institute for Particle Astrophysics and Cosmology, 452 Lomita Mall, Stanford, CA 94305, USA,}
\affiliation{5}{Canadian Institute for Theoretical Astrophysics, 60 St. George St, Toronto, ON M5S 3H8, Canada,}
\affiliation{6}{Universit\'e C\^{o}te d'Azur, Observatoire de la C\^{o}te d'Azur, CNRS, Laboratoire Lagrange, Bd de l'Observatoire, CS 34229, 06304 Nice cedex 4, France}

\correspondingauthor{Jeffersson Andres Agudelo Rueda}{jeffersson.agudelo@northumbria.ac.uk}




\begin{keypoints}


\item We characterise and determine the effect of anomalous electric fields from collisionless origin on their resolved counterparts.

\item Accurate Sub-Grid-Scale models of anomalous electric fields should consider the phase alignment with the resolved fields.

\item A phase alignment analysis reveals the impact of anomalous electric fields on the total resolved electric and magnetic fields.









\end{keypoints}

\begin{abstract}

In both collisional and collisionless plasmas the presence of a broad range of electromagnetic and plasma fluctuations provides anomalous electric fields that can be important for the dynamical evolution of the system as it is the case of magnetic reconnection, plasma turbulence and dynamo theory. In the context of plasma turbulence at scales larger than the ion's inertial length, the plasma satisfies the frozen-in condition, and the anomalous electric fields are produced by correlations between turbulent velocity and magnetic fields. Conversely, for collisionless plasmas and at kinetic scales, the total electric field has additional contributions that arise from kinetic phenomena, namely, charge separation, ambipolar electric fields and electron inertia, and each of these terms presents an anomalous counterpart. In this work we characterize the anomalous electric fields. We present a framework that can explain partial balance between the anomalous resistivity and anomalous transport, and we use it to study anomalous electric fields at kinetic scales. We establish how the different contributions to the anomalous electric field couple to the large-scale electric fields and show that the anomalous terms act back-reacting on magnetic-field organization and they may contribute to effective turbulent magnetic diffusion or shielding. Finally, we test two Sub-Grid-Scale models including anisotropic transport coefficients and show that although these models partially recover the spectral information, the phase coherence is not entirely recovered by these models suggesting a more complex non-linear contribution of the anomalous terms to the resolved scales.

\end{abstract}

\section*{Plain Language Summary}

Electric fields play a central role in controlling how energy is transferred and converted in plasmas, including those found in near-Earth space like the Earth’s magnetosheath. Turbulence generates fluctuations in plasma flows and magnetic fields that produce additional, or anomalous, electric fields beyond those predicted by simple plasma models. At very small scales, where the motion of individual charged particles becomes important, several physical processes contribute to these anomalous electric fields, however their relative importance and interaction with larger-scale plasma dynamics remain poorly understood. In this study, we investigate the origin and behaviour of anomalous electric fields in collisionless plasmas using a theoretical framework and numerical simulations. We examine how different small-scale processes contribute and interact with electric fields at larger scales. We find that the curl of the anomalous terms tend to reduce the large-scale magnetic field. We evaluate two approaches for representing unresolved small-scale physics in large-scale simulations. Although these approaches reproduce some statistical properties of the electric field, they fail to accurately capture its detailed spatial structure. These results improve our understanding of how turbulence influences energy transfer in space plasmas and provide guidance for developing more accurate large-scale plasma models.

\section{Introduction} 
\label{sec:intro}


In astrophysical and laboratory plasmas there is a large separation between the scales at which macroscopic fluid dynamics control the evolution of the system and the scales at which the plasma particles resonate with the electromagnetic fields and energy dissipation occurs. This is the case of plasma turbulence, a phenomenon in which the energy injection scale is many orders of magnitude greater than the scales at which the energy is effectively dissipated. For instance, the scale separation between injection and dissipation in the solar wind and in accretion disks is $\sim10^{6}$ \cite{bruno2013solar}, whereas for astrophysical systems, such as the intracluster-medium, the scale separation is $\sim10^{10}$ \cite{howes2024fundamental}. The scale separation poses a challenge for theoretical and numerical modeling which requires resolving the full range of length scales in the system due to the complex interplay between turbulent energy transfer across the scales and the phenomena responsible for energy conversion/dissipation at small scales. Understanding how the small-scale turbulent dynamics couple into and influence the large-scale behavior of the system and how that affects the energy budget and energy transport at system scales remains an open challenge in plasma turbulence research.

To tackle these questions a variety of frameworks have been proposed to numerically model plasma turbulence. The straightforward scenario is solving the equations of motion across all relevant scales using Direct Numerical Simulations (DNS); however, despite the vast computational power currently available, DNS of plasma turbulence are still limited to simulation domains with scale separation $\leq 10^{4}$ even in the simplified framework of magnetohydrodynamics (MHD), which does not include the full range of collisionless effects present at small scales \cite{dong2022reconnection}. To overcome this limitation of the DNS approach, some approximations include: \emph{i}) the Wentzel–Kramers–Brillouin (WKB) approximation in which the evolution of linear fluctuations on top of an inhomogeneous background are solved assuming a large scale separation between the background gradients and fluctuations \cite{parker1965dynamical,belcher1971alfvenic,alazraki1971solar, hollweg1973alfven,usmanov1996global}; \emph{ii}) turbulence transport models in which the equations are ensemble averaged in a manner similar to Reynolds-Averaged Navier-Stokes [RANS] \cite{reynolds1895iv} retaining the nonlinear terms and a coupled set of equations are evolved for both, large-scale ensemble average fields and key correlations (e.g., mean fluctuation amplitude, cross helicity, etc.) related to a given model for the averaged fluctuations \cite{marsch1989dynamics,zhou1989non,zhou1990transport,matthaeus1994evolution,zank1996evolution,zank2011transport,zank2017theory,usmanov2009mhd,usmanov2016four,usmanov2025unified,yokoi2011modeling,chhiber2021large,wang2022conservation}; \emph{iii}) mean field dynamo models in which, again, the equations are ensemble averaged and models are employed for the behavior of the averaged nonlinear terms, with particular focus on the effects that lead to the growth of the mean magnetic field \cite{brandenburg2001inverse,brandenburg2018advances,brandenburg2005astrophysical,charbonneau2010dynamo,charbonneau2020dynamo,yokoi2016new,yokoi2023unappreciated}; and \emph{iv}) large-eddy simulations (LES) in which the equations are averaged at a given scale (i.e., low-pass filtered) and ``sub-grid-scale'' (SGS) models are employed to describe the behavior of the averaged nonlinear terms \cite{shimomura1991large,smagorinsky1993,mason1994large,piomelli1999large,chernyshov2006large,chernyshov2007development,zhiyin2015large}. 

The latter three frameworks all involve averaging the dynamical equations, which, in the fully nonlinear case, results in nonlinear correlations between the unresolved fluctuations that are being averaged over, sometimes referred to as Reynolds stresses and turbulent electromotive forces, appearing in the equations for the averaged fields. Making use of these reduced frameworks requires developing appropriate models for the behavior of the nonlinear correlations, with some examples being turbulent ``eddy'' viscosity and resistivity where the Reynolds stresses or turbulent electromotive forces are modeled by a turbulent viscosity coefficient \cite{sarghini1999scale,piomelli1999large} or ``alpha'' and ``omega'' dynamo effects in the turbulent electromotive force \cite{yokoi2016new,yokoi2023unappreciated}. 

One of the main differences between the LES and turbulent transport frameworks is related to the type of average performed, with LES involving an appropriately designed weighted spatial average over a given scale that separates the dynamics into large-scale and small-scale fluctuations and turbulent transport models making use of ensemble averages (i.e., averages over many realizations of the system with small perturbations applied) that divide the dynamics into a mean background state and ``random'' turbulent fluctuations. These two approaches are conceptually similar but nuances in the properties of these averaging procedures can potentially lead to physical differences in interpretation. Although these frameworks are computationally less expensive than DNS, since they focus on the feedback of turbulence on the large-scale/background state, it is rather difficult for RANS models to capture nonlinear properties of non-steady turbulence and these models are limited by the extent to which the assumed models for the turbulent dynamics are an accurate representation of reality. 

In this work, we focus on the LES framework for examining how small-scale turbulent effects impact the evolution of the system. For LES, plasma quantities are decomposed as $f = \mean{f} + f'$ where $\mean{f}$ and $f'$ are the filtered and residual contributions respectively, and $\mean{\cdots}$ represents a suitable filter. The filtered quantities are evolved through equations of motion that depend on anomalous terms,  which are spatial averages of nonlinear products between residual terms. For example, in ideal incompressible MHD where the electric field ($\vec{E}$), magnetic field ($\vec{B}$) and plasma velocity ($\vec{u}$) satisfy ($\vec{E} + \vec{u} \times \vec{B} = 0$), the induction equation for the filtered magnetic field

\begin{eqnarray}
    \frac{\partial \mean{\vec{B}}}{ \partial t} - \nabla \times  \mean{\vec{u}} \times \mean{\vec{B}}  =  \nabla \times \vec{E}_{M},   
\end{eqnarray}

\noindent has an additional anomalous term associated with the turbulent electromotive force $\vec{E}_{M} = \mean{\vec{u} \times \vec{B}} - \mean{\vec{u}} \times \mean{\vec{B}}$ which would then be modeled in terms of filtered quantities and incorporated into the evolution of filtered quantities.  

Two of the most common approaches for modeling the SGS-scale effects on the large-scales, are the so-called functional and structural models \cite{sagaut2006large,guan2023learning}. In functional models the effect of small-scale fluctuations on the large-scales is represented by the energetic action associated with a forward energy cascade towards smaller scales and the dissipation of kinetic energy by a SGS stress tensor \cite{smagorinsky1993}. In structural models the focus is on approximating the actual structure (direction and deviatoric part) of the SGS stress tensor instead of its dissipative role \cite{leonard1975energy}. In hydrodynamic turbulence, functional SGS models rely on assumptions such as the presence of an isotropic energy cascade and thermal equilibrium of the smallest scales to parameterize the anomalous contributions as turbulent eddy viscosity \cite{sarghini1999scale}. However, these models overestimate the role of turbulent dissipation and have problems near hard boundaries or in the presence of mean flows that break the isotropy. Likewise, these assumptions do not hold in the presence of a background magnetic field or in collisionless plasmas. Alternatively, structural models rely on scale-similarity in which the contribution from the anomalous terms is thought to behave statistically similar to the smallest resolved scales; however, these models do not efficiently dissipate energy and often require the use of viscous terms \cite{bardina1980improved}. Nevertheless, SGS models have been developed to study MHD plasma turbulence, see \cite{theobald1994subgrid,agullo2001large,schmidt2011fluid,miesch2015large,aluie2017coarse,alexakis2022local}, and references therein and this has been extended to include cross helicity \cite{pouquet2022helical,yokoi2011modeling,yokoi2023unappreciated} and Hall MHD effects \cite{miura2022sub,miura2023numerical}. 

Including the effect of kinetic phenomena requires the use of simulation methods that retain the kinetic nature of the particles \cite{valentini2007hybrid,dawson1983particle}. However, these methods are computationally more expensive compared to fluid simulation methods and limit even further the size of the simulation domain. Given the additional complexity that kinetic-scale effects introduce, SGS models have typically focused on the MHD-scale nonlinear effects. However, there have been attempts to include ion kinetic effects using space-filter techniques \cite{camporeale2018coherent,cerri2020space}.

In this work, we examine the behavior of anomalous electric fields in DNS of collisionless plasma turbulence and assess the role of collisionless effects in SGS models of plasma turbulence. We focus on the anomalous contributions to generalized Ohm's law and explore possible SGS models for the anomalous contribution using three-dimensional kinetic hybrid-Vlasov simulations of plasma turbulence in the Earth's magnetosheath \cite{granier2024electron}. In these simulations ions are treated kinetically by solving the Vlasov equation for the ion distribution function in a six-dimensional phase space (3D3V, i.e., 3 spatial dimensions and 3 velocity directions), while electrons are described as a fluid. The coupling between ions and electrons is achieved through a generalized Ohm’s law that explicitly retains electron inertia capturing kinetic ion effects together with electron-inertia-driven processes, extending beyond purely ion-scale dynamics. Although many space and astrophysical plasmas have large scale separations and, thus, well developed inertial ranges, systems such as Earth's magnetosheath have more modest scale separation and are a natural laboratory in which plasma turbulence modeling can be explored combining in-situ observation and numerical simulations.

The goal of this work is not only to measure SGS electric-field terms, but to determine which physical contributions dominate them, how they couple to the resolved electric and magnetic fields, and whether simple SGS closures can reproduce them. The remainder of this manuscript is organized as follows. In section \ref{sec:data} we describe the simulation data that we analyze. In section \ref{sec:metho} we describe the theoretical framework that we use to compute and analyze the anomalous electric fields. In section \ref{sec:results} we present our results on the relation and effect between anomalous terms and filtered electric and magnetic fields. In section \ref{sec:discussion} we discuss our results and present our conclusions.


\section{Description of 3D Hybrid-Vlasov Simulations}\label{sec:data}

In this work we analyze 3D hybrid-Vlasov simulations  \cite{granier2024electron} of freely decaying turbulence with plasma and fluctuation properties mimicking those found in Earth's magnetosheath. The turbulence is initialized with magnetic fluctuations injected at scales close to the ion kinetic scales. This choice is motivated by conditions downstream of the Earth’s bow shock, where shocks and associated instabilities are expected to inject fluctuation energy directly at ion scales rather than at large fluid scales. In such environments, the bow shock effectively acts as a small-scale energy injector, producing turbulence with short correlation lengths and large fluctuation amplitudes. 

The simulations were performed using the hybrid-Vlasov-Maxwell code [HVM] \cite{valentini2007hybrid} including electron inertia in Ohm's law and an isothermal closure for the electrons. The ion to electron mass ratio $m_{i}/m_{e}=100$, where $m_{i}$ and $m_{e}$ are the ion and electron masses respectively. 
The initialization involves a superposition of 3D isotropic magnetic field fluctuations $\delta \vec{B} = \delta \vec{B}_{\perp} + \delta B_{\parallel}\hat{z}$, relative to the background magnetic field $\vec{B}_{0} = B_{0}\hat{z}$, in the wave number range $0.3 \leq kd_{i} \leq 1$ with root-mean-square (rms) amplitude $\delta B_{rms}/B_{0} \approx 0.5$, where $B_{0}$ is the amplitude of the background magnetic field, $d_{i} = c/\omega_{pi}$ is the ion inertial length, $c$ is the speed of light and $\omega_{pi}$ is the ion plasma frequency. The simulation domain is a box of size $L = 6\pi d_{i}$ with $256^{3}$ grid points. 

We analyze simulations for three values of the ion plasma beta $\beta_{i} = 0.25$, $\beta_{i} = 1$ and $\beta_{i} = 4$, and three corresponding initial ion to electron temperature ratios $T_{i0}/T_{e0} = 2.5$, $T_{i0}/T_{e0} = 10$ and $T_{i0}/T_{e0} = 40$, where $\beta_{i} = 2\mu_{0}p_{i0}/B_{0}^{2}$, $p_{i0}$ is the background scalar ion pressure, $T_{i0}$ and $T_{e0}$ are the ion and electron background temperatures. These conditions correspond to keeping the electron beta $\beta_{e}=0.1$ across the three ion beta cases ensuring a negligible finite-Larmor-radius effect ($\rho_{e}<d_{e}$), where $\beta_{e} = 2\mu_{0}p_{e0}/B_{0}^{2}$, $p_{e0}$ is the background scalar electron pressure, $d_{e}=c/\omega_{pe}$ is the electron inertial length, $\omega_{pe}$ is the electron plasma frequency, and $\rho_{e} = \sqrt{\beta_{e}}d_{e}$ is the electron gyro-radius. 

From these simulations \citeA{granier2024electron} find a shift from turbulence dominated by kinetic Alfvén Waves (KAW) to turbulence dominated by inertial KAW and inertial whistler waves at kinetic scales as well as an increase in the ion-electron decoupling for large ion beta cases. \citeA{granier2024electron} observe the development of electron-only reconnection and ion-to-electron heating ratios consistent with in-situ measurements and suggest the prevalence of electron-only magnetic reconnection events is associated with the energy injection scale. The electromagnetic spectra and the transition to electron-scale wave modes are qualitatively consistent with magnetosheath observations.

\section{Anomalous Electric Fields in Collisionless Plasmas} \label{sec:metho}

In collisionless plasmas, the electric field is given by generalized Ohm's law, such that 

\begin{eqnarray}
    {\vec{E}} = - {\vec{u}_{i}} \times {\vec{B}}  
    + \frac{{\vec{J}} \times {\vec{B}}}{{n}e}  
    + \frac{m_{e}}{m_{i}ne} \nabla \cdot {\tensnd{P}}_{i}
    - \frac{1}{ne} \nabla \cdot {\tensnd{P}}_{e}
    + \frac{m_{e}}{ne^{2}} \nabla \cdot \left({\vec{u}}_{i}{\vec{J}} + {\vec{J}}{\vec{u}}_{i} - \frac{{\vec{J}} \ {\vec{J}}}{{n}e} \right) 
    + \frac{m_{e}}{ne^2}\frac{\partial {\vec{J}}}{\partial t},  
    \label{eqn:ohm_filterd_0_3}
\end{eqnarray}

\noindent where $\vec{u}_{i}$ is the ion bulk velocity, $\vec{J}$ is the electric current density, $\tensnd{P}_{i}$ and $\tensnd{P}_{e}$ are the ion and electron pressure tensors, $n$ is the plasma density assuming quasi-neutrality $n_{i}=n_{e}=n$, and $e$ is the elementary charge. The ion pressure term can be neglected, such that

\begin{eqnarray}
\frac{m_{e}}{m_{i}ne} \nabla \cdot {\tensnd{P}}_{i}  \approx 0,    
\end{eqnarray}
\noindent as long as $m_{i} \gg m_{e}$. Neglecting the displacement current in Ampère's Law, the partial time derivative of the current is given by

\begin{eqnarray}
    \frac{\partial \vec{J}}{\partial t} = \frac{\nabla^{2} \vec{E}}{\mu_{0}},
\end{eqnarray}

\noindent where $\mu_{0}$ is the vacuum magnetic permeability, and Eq. (\ref{eqn:ohm_filterd_0_3}) becomes

\begin{eqnarray}
    {\vec{E}} = - {\vec{u}_{i}} \times {\vec{B}}  
    + \frac{{\vec{J}} \times {\vec{B}}}{{n}e}
    - \frac{1}{ne} \nabla \cdot {\tensnd{P}_{e}}  
    + \frac{m_{e}}{ne^{2}} \left[ \nabla \cdot \left({\vec{u}}_{i}{\vec{J}} + {\vec{J}}{\vec{u}}_{i} - \frac{{\vec{J}} \ {\vec{J}}}{{n}e} \right) 
    + \frac{1}{\mu_{0}}\nabla^{2}\vec{E} \right],  
    \label{eqn:ohm_filterd_0_4}
\end{eqnarray}

\noindent which, with a further approximation of constant density in front of the Laplacian of $\vec{E}$ ($\nabla^{2}\vec{E}/n\mu_{0} \approx \nabla^{2}\vec{E}/n_{0}\mu_{0}$), is the form used by the simulations analyzed in this study. For simplicity we refer to the MHD term $\vec{E}_{MHD} = \vec{u}_{i} \times \vec{B}$, the Hall term $\vec{E}_{Hall} = \vec{J} \times \vec{B}/ne$, the ambipolar/diamagnetic term associated with the divergence of the electron pressure $\vec{E}_{P_{e}} = \nabla \cdot \tensnd{P}_{e}/ne$  and the electron inertia term $\vec{E}_{Ine} = m_{e} (\nabla \cdot (\vec{u}_{i}\vec{J} + \vec{J}\vec{u}_{i} - \vec{J}\vec{J}/ne) + \nabla^{2}\vec{E}\mu_{0})/ne^2$. Note that since $-\vec{E}_{MHD}$ and $\vec{E}_{Hall}$ act at different scales, we retain their separate contributions to the electric field instead of combining $\vec{E}_{MHD}$ and $\vec{E}_{Hall}$ in the single term $\vec{E}_{EMHD} = \vec{E}_{MHD} - \vec{E}_{Hall} = \vec{u}_{e} \times \vec{B} $, where $\vec{u}_{e}$ is the electron bulk velocity.

For our analysis, we apply a Gaussian filter to Eq. (\ref{eqn:ohm_filterd_0_4}) such that a filtered quantity is given by 

\begin{eqnarray}
    \mean{f}(\vec{x}) = \frac{1}{\Delta V} \int f(\vec{y})G_{l}(\vec{x} - \vec{y})d^3{\bf y},
    \label{eqn:convo}
\end{eqnarray}

\noindent where

\begin{eqnarray}
    G_{l}(\vec{x}) = \left( \frac{1}{2\pi\Delta_{l}^{2}} \right)^{3/2}\exp\left(-\frac{\vec{x}^{2}}{2\Delta_{l}^{2}}\right),
    \label{eqn:filter}
\end{eqnarray}

\noindent and $\Delta_{l}^2$ is the variance associated with the filter kernel. We take the cutoff scale associated with the filter to be $k_{c} = \pi/\Delta_{c}$, where $\Delta_{c}=\sqrt{12}\Delta_{l}$ \cite{leonard1975energy,pope2001turbulent}. By decomposing quantities as $f=\mean{f}+f^\prime$, the filtered version of Eq.(\ref{eqn:ohm_filterd_0_4}) becomes 

\begin{eqnarray}
    \mean{\vec{E}} = &-& \mean{\vec{u}_{i}} \times \mean{\vec{B}}  
    + \frac{\mean{\vec{J}} \times \mean{\vec{B}}}{\mean{n}e}
    - \frac{1}{\mean{n}e} \nabla \cdot \mean{\tensnd{P}_{e}}  
    + \frac{m_{e}}{\mean{n}e^{2}} \left[ \nabla \cdot \left(\mean{\vec{u}}_{i}\mean{\vec{J}} + \mean{\vec{J}}\mean{\vec{u}}_{i} - \frac{\mean{\vec{J}} \ \mean{\vec{J}}}{\mean{n}e} \right) + \frac{1}{\mu_{0}}\nabla^{2}\mean{\vec{E}} \right]  \nonumber \\
    &-& \boldsymbol{\tau}_{nE}
    - \boldsymbol{\tau}_{nu_{i} \times B}  
    + \boldsymbol{\tau}_{J \times B}   
    +  \boldsymbol{\tau}_{Ine},
    \label{eqn:ohm_filterd_3}
\end{eqnarray}

\noindent where

\begin{eqnarray}
    \boldsymbol{\tau}_{nE} &=& (\mean{n\vec{E}} - \mean{n}\mean{\vec{E}})/\mean{n}, \\
    \boldsymbol{\tau}_{nu_{i} \times B} &=& (\mean{n \vec{u}_{i} \times \vec{B}} -\mean{n} \ \mean{\vec{u}}_{i}\times\mean{\vec{B}})/\mean{n}, \\
     \boldsymbol{\tau}_{J \times B} &=& (\mean{\vec{J} \times \vec{B}} - \mean{\vec{J}} \times \mean{\vec{B}}) /\mean{n}e,  \\
    \boldsymbol{\tau}_{Ine} &=&  \frac{m_{e}}{\mean{n}e^{2}}\nabla \cdot \left[ \mean{  \left(\vec{u_{i}}\vec{J} + \vec{J}\vec{u_{i}} - \frac{\vec{J} \vec{J}}{ne} \right)} - \left(\mean{\vec{u}}_{i}\mean{\vec{J}} + \mean{\vec{J}}\mean{\vec{u}}_{i} - \frac{\mean{\vec{J}} \ \mean{\vec{J}}}{\mean{n}e} \right) \right],
\end{eqnarray}

\noindent are the anomalous contributions to the filtered electric field arising from the combination of nonlinear interactions between residual fluctuations and themselves, and between residual fluctuations and the filtered fluctuations. In Eq. (\ref{eqn:ohm_filterd_3}), the term $\mean{\vec{E}}$ is the filtered electric field, $\mean{\vec{E}}_{MHD} = \mean{\vec{u}_{i}} \times \mean{\vec{B}}$ is the MHD term based only on filtered fields. Likewise, $\mean{\vec{E}}_{Hall} = \frac{\mean{\vec{J}} \times \mean{\vec{B}}}{\mean{n}e}$ (Hall term), $ \mean{\vec{E}}_{P_{e}} =  \frac{1}{\mean{n}e} \nabla \cdot \mean{\tensnd{P}_{e}}$ (ambipolar/diamagnetic term) and $\mean{\vec{E}}_{Ine} = \frac{m_{e}}{\mean{n}e^{2}} \left[ \nabla \cdot \left(\mean{\vec{u}}_{i}\mean{\vec{J}} + \mean{\vec{J}}\mean{\vec{u}}_{i} - \frac{\mean{\vec{J}} \ \mean{\vec{J}}}{\mean{n}e} \right) + \frac{1}{\mu_{0}}\nabla^{2}\mean{\vec{E}} \right] $ (electron inertia term) are terms based on filtered fields. Note that since $\mean{\vec{u}_{i} \times \vec{B}} \neq \mean{\vec{E}}_{MHD}$, $\mean{\vec{J} \times \vec{B}/ne} \neq \mean{\vec{E}}_{Hall}$, and similar for the inertia term, $\mean{\vec{E}}_{MHD}$, $\mean{\vec{E}}_{Hall}$, $\mean{\vec{E}}_{P_{e}}$, and $\mean{\vec{E}}_{Ine}$ are the contributions to filtered generalized Ohm's law Eq. (\ref{eqn:ohm_filterd_3}). For simplicity and without ambiguity in what follows we refer to $\mean{\vec{E}}, \mean{\vec{E}}_{MHD}, \mean{\vec{E}}_{Hall}, \mean{\vec{E}}_{P_{e}}$ and $\mean{\vec{E}}_{Ine}$ as filtered quantities. We also refer to $\boldsymbol{\tau}_{nE}$ as the anomalous field-density correlation term, $\boldsymbol{\tau}_{nu_{i} \times B}$ as the anomalous MHD term, $\boldsymbol{\tau}_{J \times B}$ as the anomalous Hall term and $\boldsymbol{\tau}_{Ine}$ as the anomalous electron inertia term. Note that the anomalous contributions contain information from both the residual fluctuations and their interaction with the filtered quantities. Note as well that neither $\nabla \cdot \tensnd{P}_{e}/ne$ nor $\nabla ^{2}\vec{E}/\mu_{0}$ explicitly have anomalous contributions in the above formulation since $\mean{\nabla \cdot \tensnd{P}_{e}} = \nabla \cdot \mean{\tensnd{P}_{e}}$ and $\mean{\nabla^{2} \vec{E}} = \nabla \cdot \mean{\vec{E}}$. However, $\boldsymbol{\tau}_{nE}$ implicitly contains the effects of the compressive nonlinearities associated with density correlating with all of the terms in generalized Ohm's law, including ${\bf E}_{P_{e}}$ and $\frac{m_e}{ne^2\mu_0}\nabla^2{\bf E}$.

Since we are also interested in addressing the combined effect of the anomalous terms, we define the total anomalous electric field as

\begin{equation}
\Sigma\boldsymbol{\tau} = -\boldsymbol{\tau}_{nE} - \boldsymbol{\tau}_{nu_i \times B} + \boldsymbol{\tau}_{J \times B} + \boldsymbol{\tau}_{Ine}.
\end{equation}

Furthermore, from the perspective that in the context of fluid (MHD or Hall MHD) simulations, many of the filtered terms in Eq. (\ref{eqn:ohm_filterd_3}) are also effectively anomalous physics, we define the net contributions
\begin{equation}
\Sigma\boldsymbol{\tau}_{MHD} = \mean{\vec{E}}_{Hall} - \mean{\vec{E}}_{P_{e}} + \mean{\vec{E}}_{Ine} + \Sigma\boldsymbol{\tau},
\label{eqn:tau_MHD}
\end{equation}

\noindent and 

\begin{equation}
\Sigma\boldsymbol{\tau}_{Hall} = - \mean{\vec{E}}_{P_{e}} + \mean{\vec{E}}_{Ine} + \Sigma\boldsymbol{\tau},
\label{eqn:tau_Hall}
\end{equation}

and examine them through the lenses of anomalous dynamics. Note that we use ``anomalous'' in two related but distinct senses: $\Sigma \boldsymbol{\tau}$ denotes true filter-induced SGS correlations, whereas $\Sigma \boldsymbol{\tau}_{MHD}$ and $\Sigma \boldsymbol{\tau}_{Hall}$ denote all terms that would be absent from a reduced MHD or Hall-MHD closure.

\subsection{Balance between anomalous terms}
\label{sec:balance_anomalous}
It has been reported that the anomalous resistivity ($\boldsymbol{\tau}_{nE}$) and the anomalous transport at electron scales ($\boldsymbol{\tau}_{nu_{e} \times B} = \mean{n\vec{u}_{e} \times \vec{B}} - \mean{n} \mean{\vec{u}_{e}} \times \mean{\vec{B}}$) due to the presence of Lower Hybrid waves (LHW) show partial balance providing a negligible net contribution to the reconnecting electric field \cite{graham2022direct,zhong2025electromagnetic}. \citeA{graham2022direct} argues that the mismatch in the balance between $\boldsymbol{\tau}_{nE}$ and $\boldsymbol{\tau}_{nu_{e} \times B}$ is due to measurement limitations or due to the breakdown of the electron frozen-in condition. However, the partial balance between $\boldsymbol{\tau}_{nE}$ and $\boldsymbol{\tau}_{nu_{e}\times B}$ can also be explained without breaking down the frozen-in condition. We show first the origin of a possible balance between $\boldsymbol{\tau}_{nE}$ and $\boldsymbol{\tau}_{nu_{i}\times B}$ by considering the case in which the bulk flow is carried by ions and $\vec{E}_{MHD} = -\vec{u}_{i} \times \vec{B}$. Considering that $\mean{\vec{E}_{MHD}} = - \mean{\vec{u}_{i} \times \vec{B}}$, and $\mean{n\vec{E}_{MHD}} = - \mean{n\vec{u}_{i} \times \vec{B}}$, leads to

\begin{eqnarray}
\boldsymbol{\tau}_{nE_{MHD}} &=&  \left( \mean{n} \ \mean{\vec{u}_{i}}\times \mean{\vec{B}} - \mean{n\vec{u}_{i}\times \vec{B} } \right)/\mean{n} + \mean{\vec{u}_{i}\times \vec{B} } - \mean{\vec{u}_{i}}\times \mean{\vec{B}} \nonumber \\ 
\boldsymbol{\tau}_{nE_{MHD}} &=& -\boldsymbol{\tau}_{nu_{i}\times B} + \boldsymbol{\tau}_{u_{i} \times B},     
\end{eqnarray}

\noindent where $\boldsymbol{\tau}_{u_{i}\times B} = \mean{\vec{u}_{i}\times \vec{B} } - \mean{\vec{u}_{i}}\times \mean{\vec{B}}$ is the anomalous term associated with the nonlinear interaction between $\vec{u}_{i}$ and $\vec{B}$. Thus, in the MHD case, a perfect balance would be a consequence of the electric field being mainly dominated by the MHD term and a negligible $\boldsymbol{\tau}_{u_{i} \times B}$. As such the combination $\tau_{nE_{MHD}} + \tau_{n u_i \times B}$ removes the compressive component of the MHD nonlinearity associated with density fluctuations correlating with convective electric fields, producing a natural partial anti-correlation between these two terms.

At scales below the ion gyroradius, $\vec{E} = -\vec{E}_{EMHD} - \vec{E}_{P_{e}} + \vec{E}_{Ine}$, and the anomalous field-density correlation term $\boldsymbol{\tau}_{nE_{EMHD}} = (\mean{n\vec{E}_{EMHD}} - \mean{n} \mean{\vec{E}}_{EMHD})/\mean{n}$ satisfies 

\begin{eqnarray}
 \boldsymbol{\tau}_{nE_{EMHD}} = -\boldsymbol{\tau}_{nu_{e} \times B} + \boldsymbol{\tau}_{u_{e} \times B},   
\end{eqnarray}

\noindent which shows that an imbalance can emerge due to the presence of nonlinear correlations between $\vec{u}_{e}$ and $\vec{B}$ ($\boldsymbol{\tau}_{u_{e}\times B} \neq 0$), which can occur even in ideal electron MHD. Thus, in the context of magnetic reconnection \cite{graham2022direct,zhong2025electromagnetic,stanish2025turbulent}, at scales $l \gg d_{i}$ and $l \gg \rho_{i}$ it is expected that the anomalous contributions to the reconnecting electric field, $\boldsymbol{\tau}_{nE}$ and $\boldsymbol{\tau}_{nu_{i} \times B}$, balance each other and only $\boldsymbol{\tau}_{u_{i} \times B}$ contributes. Likewise, at smaller scales, $l \sim d_{i}$ and $l \gg \rho_{e}$, $\boldsymbol{\tau}_{nE}$ and $\boldsymbol{\tau}_{nu_{e} \times B}$ balance each other and only $\boldsymbol{\tau}_{u_{e} \times B}$ contributes. Furthermore, in regions where there are non-negligible $\vec{E}_{{P_{e}}}$ and $\vec{E}_{Ine}$ contributions that breakdown the frozen-in condition for electrons, there is an anomalous contribution from the electron inertia ($\boldsymbol{\tau}_{Ine}$) as the pressure term does not directly contribute to the anomalous electric field.

To understand at which scales the electron inertia term and its anomalous contributions can be important, we consider that, at scales at which the current is carried only by electrons ($\vec{u}_{i}\approx 0$), the inertia term can be expressed as $-(m_{e}/ne) [\nabla \cdot (n\vec{u}_{e}\vec{u}_{e}) + \partial n\vec{u}_{e}/\partial t ]$ and the generalized Ohm's law takes the form

\begin{eqnarray}
    \vec{E} = - \vec{u}_{e} \times \vec{B} - \frac{1}{ne}\nabla \cdot \tensnd{P}_{e} - \frac{m_{e}}{ne} \left[ \nabla \cdot  (n\vec{u}_{e}\vec{u}_{e}) + \frac{\partial n \vec{u}_{e}}{\partial t} \right], 
\end{eqnarray}

\noindent which shows that the only anomalous terms are $\boldsymbol{\tau}_{nE}$, $\boldsymbol{\tau}_{nu_{e}\times B}$ and 

\begin{eqnarray}
    \boldsymbol{\tau}_{Ine} \approx \frac{m_{e}}{\mean{n}e} \nabla \cdot (\mean{n \vec{u}_{e}\vec{u}_{e}} - \mean{n} \ \mean{\vec{u}_{e}\vec{u}_{e}}) + \frac{m_{e}}{\mean{n}e} \frac{\partial}{\partial t} \left( \mean{n\vec{u}_{e}} - \mean{n} \ \mean{\vec{u}_{e}}  \right).
\end{eqnarray}

Although the inertia term is small, a dimensional analysis for the ratios $\vec{u}_{e} \times \vec{B}/[\nabla \cdot nm_{e}\vec{u}_{e}\vec{u}_{e})/ne]$ and $\vec{u}_{e} \times \vec{B}/[ (m_{e}/ne)\partial (n\vec{u}_{e})/\partial t]$, shows that the inertia term can be important for time changes $\Delta t \approx \Omega_{ce}^{-1}$, where $\Omega_{ce}=eB/m_{e}$ is the electron gyrofrequency, which corresponds to spatial variations at the scale of the electron gyroradius $\rho_{e}$.


\begin{figure*}{}
\centering
\includegraphics[width=1.0\linewidth]{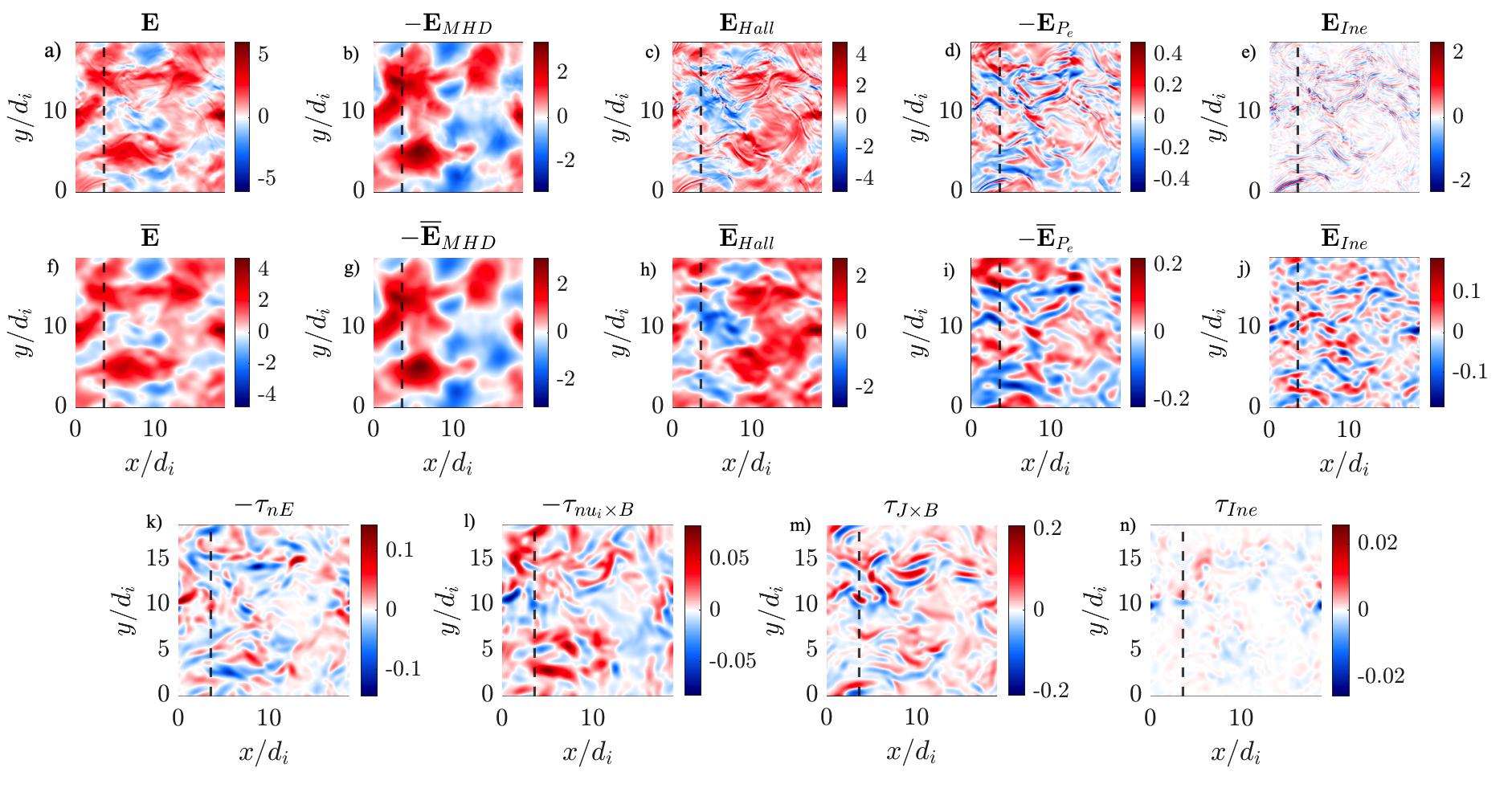}
\caption{2D cuts in the $xy$-plane of the simulation domain at $z = 3.5d_{i}$ for $\beta_{i}=0.25$ at the time of maximum dissipation. Panels (a-e) depict the x-component of $\vec{E}$, $-\vec{E}_{MHD}$, $\vec{E}_{Hall}$, $-\vec{E}_{P_{e}}$ and $\vec{E}_{Ine}$. Panels (f-j) depict the x-component of $\mean{\vec{E}}$, $-\mean{\vec{E}}_{MHD}$, $\mean{\vec{E}}_{Hall}$, $- \mean{\vec{E}}_{P_{e}}$ and  $\mean{\vec{E}}_{Ine}$. Panels (k-n) depict the x-component of $-\boldsymbol{\tau}_{nE}$, $-\boldsymbol{\tau}_{u_{i} \times B}$,  $\boldsymbol{\tau}_{J \times B}$ and $\boldsymbol{\tau}_{Ine}$. All panels are normalized to $\langle |E_{x}| \rangle$.}
\label{fig:composed_E_taus}
\end{figure*}

\begin{figure*}{}
\centering
\includegraphics[width=1.0\linewidth]{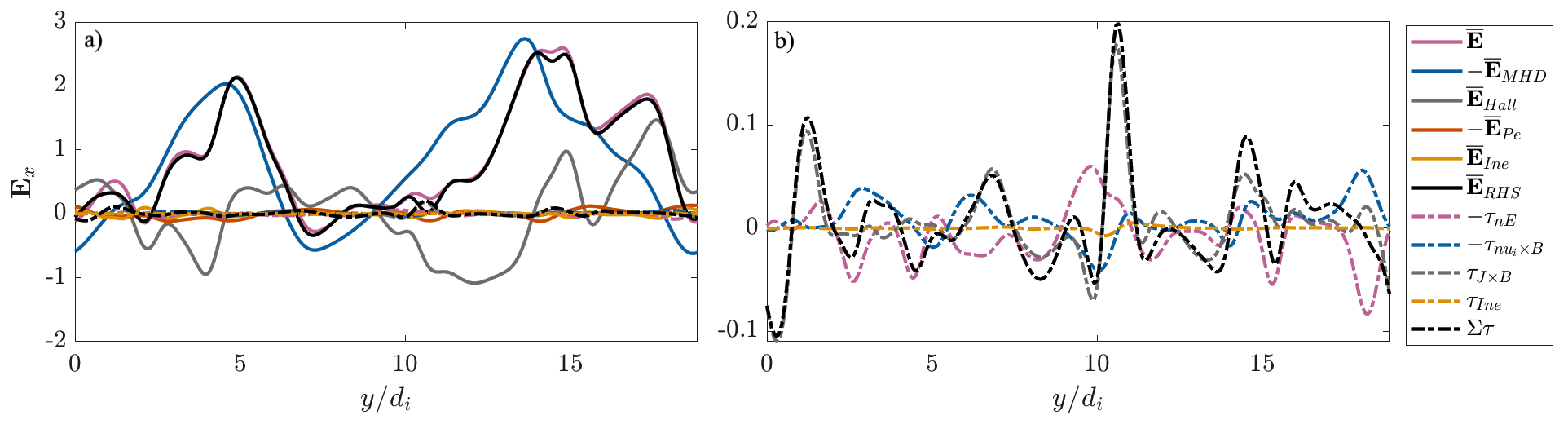}
\caption{1D cuts along the black dashed-line in the 2D cuts shown in Figure \ref{fig:composed_E_taus}. a) Filtered and anomalous terms in the filtered generalized Ohm's Law (Figure \ref{fig:composed_E_taus}(f-n)): $\mean{\vec{E}}$ (light purple solid-line), $\mean{\vec{E}}_{MHD}$ (blue solid-line), $\mean{\vec{E}}_{Hall}$ (gray solid-line), $\mean{\vec{E}}_{P_{e}}$ (vermilion solid-line), $\mean{\vec{E}}_{Ine}$ (yellow solid-line), $\boldsymbol{\tau}_{nE}$ (light purple dot-dashed-line), $\boldsymbol{\tau}_{nu_{i}\times B}$ (blue dot-dashed-line), $\boldsymbol{\tau}_{J\times B}$ (gray dot-dashed-line) and $\boldsymbol{\tau}_{Ine}$ (yellow dot-dashed-line). The black solid-line corresponds to $\mean{\vec{E}}_{RHS}$ and the black dot-dashed-line corresponds to $\Sigma \boldsymbol{\tau}$. b) Zoom in to the anomalous terms in a).}
\label{fig:1E_taus_1D_2}
\end{figure*}

\section{Results}\label{sec:results}

We first illustrate the effect of filtering fields using a filter-scale $\Delta_{c} = 0.96 d_{i}$ for the $\beta_{i}=0.25$ case. Figure \ref{fig:composed_E_taus} shows 2D cuts in the $xy$-plane of the simulation domain at $z = 3.5d_{i}$, for the x-components of the terms in Eq. (\ref{eqn:ohm_filterd_0_4}), Figure \ref{fig:composed_E_taus}(a-e), and the filtered terms in Eq. (\ref{eqn:ohm_filterd_3}), Figure \ref{fig:composed_E_taus}(f-j). For all panels blue represents negative values and red positive values and each quantity is normalized to $\langle |E_{x}| \rangle$, where $\langle ... \rangle$ is the spatial average. The electric field shows structures at several scales. Comparing Figure \ref{fig:composed_E_taus}(a-e), we observe that larger scale $\vec{E}$ structure ($\sim$5 to 10$d_{i}$) are mainly associated with $\vec{E}_{MHD}$, whereas smaller scale structures ($ < 5 d_{i}$) are associated with $\vec{E}_{Hall}$, as qualitatively expected from dimensional analysis of the terms in generalized Ohm's law. Conversely $\vec{E}_{P_{e}}$ and $\vec{E}_{Ine}$ are localized to very small structures with thickness less than $1d_{i}$.

Looking at the filtered quantities, Figure \ref{fig:composed_E_taus}(f), $\mean{\vec{E}}$ does not show dominant small-scale features as they have been removed by the filter and $\mean{\vec{E}}$ corresponds mainly to $-\mean{\vec{E}}_{MHD}$. Recalling that quantities depicted in Figure \ref{fig:composed_E_taus}(g-j) are contributions to generalized Ohm's law based purely on the information in the filtered quantities, $\mean{\vec{E}}$ corresponds mainly to a combination of $\mean{\vec{E}}_{MHD}$ and $\mean{\vec{E}}_{Hall}$, both featuring structures of similar scales and amplitudes. $\mean{\vec{E}}_{P_{e}}$ contributes to smaller scale structures slightly elongated along the x-direction in the filtered electric field, but with approximately one order of magnitude lower amplitude than $\mean{\vec{E}}_{MHD}$ or $\mean{\vec{E}}_{Hall}$. The elongation is due to the polarization of the electric field, the y-component of $\mean{\vec{E}}_{P_{e}}$ (not shown here) depicts structures elongated in the y-direction. $\mean{\vec{E}}_{Ine}$ contributes to yet smaller scales with roughly isotropic structures in the xy-plane, which, unlike $\vec{E}_{Ine}$, are significantly lower amplitude than $\mean{\vec{E}}_{MHD}$ or $\mean{\vec{E}}_{Hall}$.

Figure \ref{fig:composed_E_taus}(k-n) depict, the anomalous terms in Eq. (\ref{eqn:ohm_filterd_3}). The anomalous contributions are approximately an order of magnitude smaller than the filtered contributions. $\boldsymbol{\tau}_{nE}$, $\boldsymbol{\tau}_{nu_{i} \times B}$ and  $\boldsymbol{\tau}_{J \times B}$ show small clusters while even the most intense values of $\boldsymbol{\tau}_{Ine}$ are an order of magnitude smaller than the other anomalous terms. $\boldsymbol{\tau}_{Ine}$ spatial distribution interestingly exhibits localized hotspots of activity.  

To facilitate comparison between terms, Figure \ref{fig:1E_taus_1D_2} shows 1D cuts, along the y-direction (black dashed-line in Figure \ref{fig:composed_E_taus}), for the x-component of the filtered (solid-lines) and anomalous (dot-dashed-lines) terms of Eq. (\ref{eqn:ohm_filterd_3}). The large-scale variation of $\mean{\vec{E}}$ is provided by $-\mean{\vec{E}}_{MHD}$ and with significant deviations resulting from $\mean{\vec{E}}_{Hall}$, whereas $-\mean{\vec{E}}_{P_{e}}$ and $\mean{\vec{E}}_{Ine}$ are very small compared to the other filtered terms. The total contribution from the filtered terms on the right-hand-side of Eq. (\ref{eqn:ohm_filterd_3}), 

\begin{eqnarray}
\mean{\vec{E}}_{RHS} = -\mean{\vec{E}}_{MHD} + \mean{\vec{E}}_{Hall} - \mean{\vec{E}}_{P_{e}} + \mean{\vec{E}}_{Ine},     
\end{eqnarray}

\noindent shows small deviations from $\mean{\vec{E}}$ corresponding to localized regions of enhanced $\Sigma \boldsymbol{\tau}$. Looking at the anomalous terms, Figure \ref{fig:1E_taus_1D_2}(b) $-\boldsymbol{\tau}_{nE}$, $-\boldsymbol{\tau}_{u_{i} \times B}$ and $\boldsymbol{\tau}_{J \times B}$ have mostly similar amplitude fluctuations except for localized regions where $\boldsymbol{\tau}_{J \times B}$ is considerably larger than $\boldsymbol{\tau}_{nE}$ and $-\boldsymbol{\tau}_{u_{i} \times B}$. The terms $-\boldsymbol{\tau}_{nE}$ and $-\boldsymbol{\tau}_{u_{i} \times B}$ show regions where they roughly balance each other, such as in the vicinity of $y/d_{i}=6$ or $y/d_{i}=10$; however, this behavior is not always the case. In these regions, $\Sigma \boldsymbol{\tau} $ is mainly sustained by $ \boldsymbol{\tau}_{J \times B}$.

\begin{figure*}{}
\centering
\includegraphics[width=1.0\linewidth]{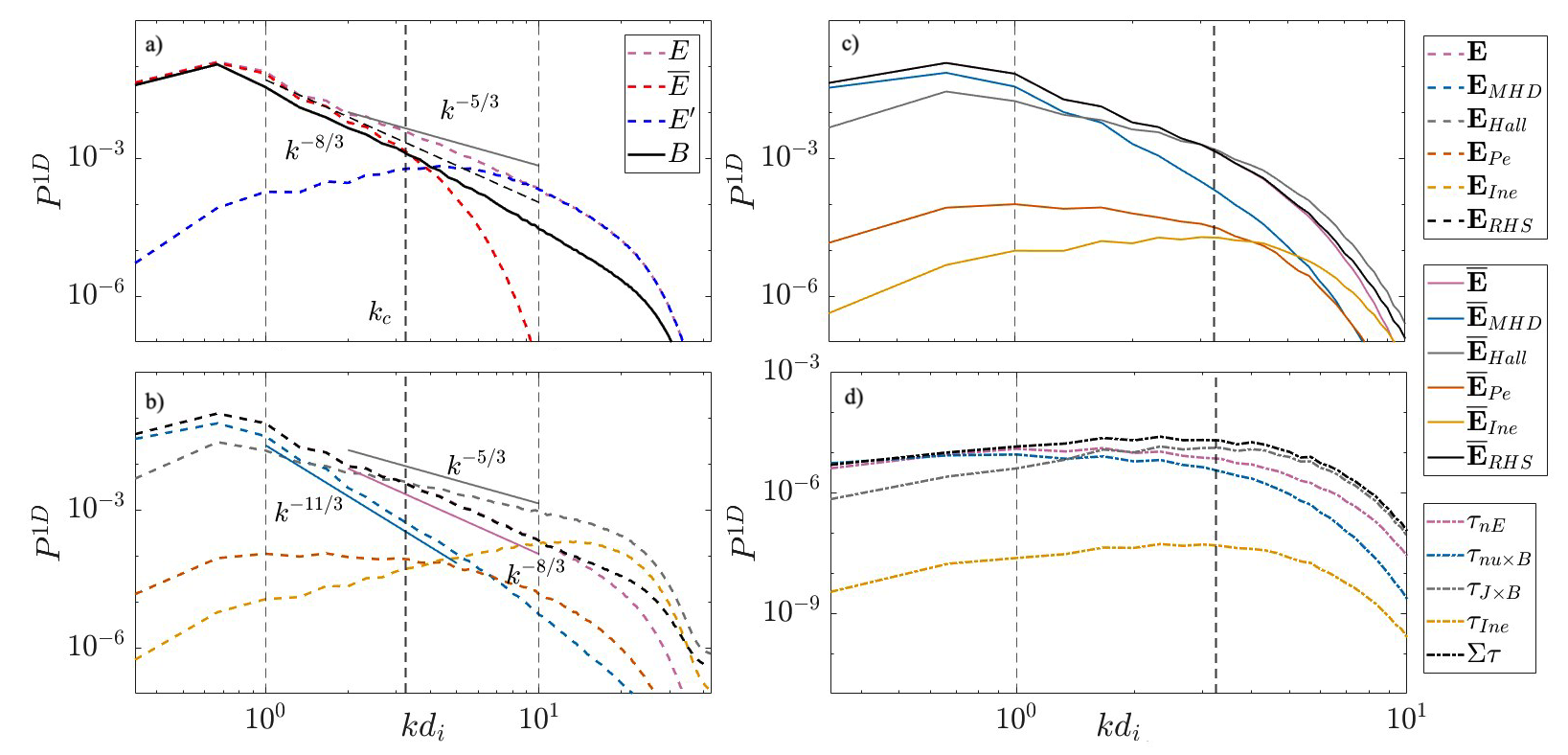}
\caption{Spectral information for $\Delta_{c} =0.96 d_{i}$ and $\beta_{i}=0.25$. a) Omnidirectional power spectral density of the magnetic field, $P^{1D}_{\vec{B}}$ (black solid-line); unfiltered electric field, $P^{1D}_{\vec{E}}$, (light-purple dashed-line); filtered electric field, $P^{1D}_{\mean{\vec{E}}}$, (red dashed-line) and residual electric field $P^{1D}_{\vec{E}'}$ (blue dashed-line). b-c) $P^{1D}$ of unfiltered electric field terms (dashed-lines in panel b), filtered terms (solid lines in panel c) and anomalous terms (dot-dashed-lines in panel d). The color convention is the same as in Figure \ref{fig:1E_taus_1D_2}. The thick vertical dashed-line represents the filter cutoff-scale ($k_{c}$), the thin dashed-lines mark $kd_{i}=1$ and $kd_{e}=1$ and spectral scalings are included for reference.}
\label{fig:1spectrum_E_taus}
\end{figure*}

 \subsection{Spectral Analysis of the Filtered and Anomalous Terms}
 \label{sec:spectrum_ana}
In addition to understanding the effect of filtering in real space, we explore the spectral information associated with the unfiltered, filtered and anomalous contributions to generalized Ohm's law. We compute, for any vector quantity $\vec{E}$, the omnidirectional power spectral density 

\begin{eqnarray}
P^{1D}_{\vec{E}}(k) = \sum_{k \in shell } \sum_{\eta=1}^{3}\frac{ \delta\tilde{E}_{\eta} \delta\tilde{E}^{*}_{\eta}} {\Delta k},
\end{eqnarray}

\noindent where $\delta E_{\eta} = E_{\eta} - \langle E_{\eta} \rangle$ is the $\eta_{th}$ component of the fluctuation $\delta \vec{E}$, $\delta \tilde{E}_{\eta}$ is its Fourier's transform and $\delta \tilde{E}_{\eta}^{*}$ is the complex conjugate of $\delta \tilde{E}_{\eta}$ and  $k=\sqrt{k_{x}^{2} + k_{y}^{2} + k_{z}^{2}}$ is the magnitude of the wave vector. 

To illustrate the effect of filtering in k-space, Figure \ref{fig:1spectrum_E_taus}(a) shows the omnidirectional power spectral density of the unfiltered magnetic field ($P^{1D}_{\vec{B}}$), unfiltered electric field ($P^{1D}_{\vec{E}}$), filtered electric field ($P^{1D}_{\mean{\vec{E}}}$) and residual electric field ($P^{1D}_{\vec{E}'}$). The solid gray-line shows the scaling $\sim k^{-5/3}$ (for reference only). $P^{1D}_{\vec{B}}$ follows a scaling $\sim k^{-8/3}$ for $kd_{i}<4$ which is typical in the kinetic range. Likewise the electric field, $P^{1D}_{\vec{E}}$, follows $\sim k^{-8/3}$ for $kd_{i}<2$ and slightly flattens towards $\sim k^{-5/3}$ \cite{granier2024electron}. $P^{1D}_{\mean{\vec{E}}}$ follows $\sim k^{-8/3}$ for $kd_{i}<1$ and then exponentially decays at larger $k$ as a consequence of the Gaussian filter. $P^{1D}_{\vec{E}'}$ intercepts with $P^{1D}_{\mean{\vec{E}}}$ near the cutoff scale (vertical black dashed-line), $k_{c}=3.23$ for $\Delta_{c}=0.96 d_{i}$ and $\beta_{i}=0.25$. The spectral energy below the cutoff scale is stored in the the residual part.   

Since different terms in the generalized Ohm's law act at different scales, it is expected that different spectral behavior will be obtained in different scale ranges. In Figure \ref{fig:1spectrum_E_taus}(b), the spectra of each individual term in Ohm's law is plotted. The spectrum of the unfiltered MHD term $P^{1D}_{\vec{E}_{MHD}}$ follows a slope slightly steeper than $\sim k^{-11/3}$, and the unfiltered Hall term, $P^{1D}_{\vec{E}_{Hall}}$, follows a slope shallower than $\sim k^{-5/3}$. At scales $1<kd_{i}<10$, $P^{1D}_{{\vec{E}_{Hall}}}$ is shallower than $P^{1D}_{{\vec{E}}_{MHD}}$ and around $k\rho_{i}=1$ ($kd_{i}=2$) there is a transition and the Hall term carries most of the power \cite{stawarz2021comparative,lewis2023magnetospheric}. 

The spectrum of the ambipolar/diamagnetic term, $P^{1D}_{\vec{E}_{P_{e}}}$, is flat for $kd_{i}<4$ which corresponds to twice the gyro-radius-scale $k\rho_{i}$ and then it steepens. However, the effect of the filtering employed to provide dissipation of the fluctuations energy at large $k$ \cite{granier2024electron} obscures the estimation of any spectral slope at these scale. The spectrum of the inertia term $P^{1D}_{\vec{E}_{In}}$ has a positive slope $\sim k^{4/3}$, the spectral energy increases towards smaller scales in the kinetic range and it peaks at $kd_{i}=10.2$ (i.e., $kd_{e} \sim 1$). All terms follow the exponential decay around $kd_{i} = 22$ associated with the aforementioned energy dissipation method. At $k<k_{c}$, all filtered terms follow a similar scaling as their unfiltered counterpart, see Figure \ref{fig:1spectrum_E_taus}(c). For $k>k_{c}$ the spectra for all the filtered contributions follow an exponential decay towards $kd_{i}=10.66$ due to the Gaussian scale filtering.     

Looking at the spectra of anomalous terms, see Figure \ref{fig:1spectrum_E_taus}(d), $P^{1D}_{\boldsymbol{\tau}_{nE}}$ shows a nearly flat spectrum in the range $1<kd_{i}<k_{c}d_{i}$, followed by an exponential decay for $k>k_{c}$. Although $P^{1D}_{\boldsymbol{\tau}_{nu_{i} \times B}}$ remains constant between $1<kd_{i}<2.3$, it steepens before $k_{c}$. $P^{1D}_{\boldsymbol{\tau}_{J \times B}}$ and $P^{1D}_{\boldsymbol{\tau}_{Ine}}$ show a positive spectrum with energy increasing at smaller scales. For $kd_{i}>k_{c}d_{i}$, the spectral energy of $\boldsymbol{\tau}_{nu_{i} \times B}$ decays faster than the spectral energy of $\boldsymbol{\tau}_{nE}$ and $\boldsymbol{\tau}_{J \times B}$.

The spectrum of the total anomalous term, $P^{1D}_{\Sigma \boldsymbol{\tau}}$  shows a positive slope $1<kd_{i}<k_{c}d_{i}$ followed by an exponential decay that matches the decay of $P^{1D}_{\boldsymbol{\tau}_{J \times B}}$. The energy stored in $\Sigma \boldsymbol{\tau}$ is two orders of magnitude smaller than the spectrum of the filtered terms for $kd_{i} \lesssim 1$. This is because the amplitude of the anomalous terms is approximately a tenth of the filtered terms (Figure \ref{fig:1E_taus_1D_2}). 

\begin{figure*}{}
\centering
\includegraphics[width=0.7\linewidth]{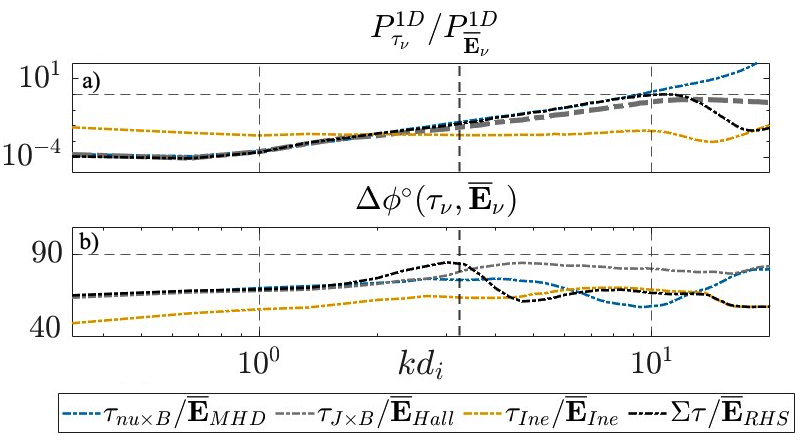}
\caption{Spectral ratios: a) Ratio of the power spectra for the anomalous and filtered terms as a function of $k$. The horizontal dashed-line marks $10^{0}$. b) Phase difference ($\Delta \phi$) between the anomalous and filtered terms as a function of $k$. The horizontal dashed-line marks $90^{\circ}$. The color convention in both panels is the same as in Figure \ref{fig:1E_taus_1D_2}. The thick vertical dashed-line marks the cut-off scale $k_{c}$ and the thin vertical dashed-lines at $kd_{i}=1$ and $k_{di}=10$ ($k_{de}=1$) mark the ion and electron scales respectively.  }
\label{fig:1spectrum_E_taus2}
\end{figure*}

By construction the anomalous and filtered terms are linked, e.g., in light of $\mean{\vec{u}_{i}\times\vec{B}} =  \mean{\vec{E}}_{MHD} + \boldsymbol{\tau}_{nu_{i} \times B}$. Although their relative amplitude can be intuitive, their level of alignment characterized by the complex phase shift between anomalous and filtered terms is less clear. To explore the spectral relation between the filtered and anomalous terms as well as their contributions to the filtered electric field, we compute the spectral ratios $P^{1D}_{\boldsymbol{\tau}_{nu_{i}\times B}}/P^{1D}_{\mean{\vec{E}}_{MHD}}$, $P^{1D}_{\boldsymbol{\tau}_{J\times B}}/P^{1D}_{\mean{\vec{E}}_{Hall}}$, $P^{1D}_{\boldsymbol{\tau}_{Ine}}/P^{1D}_{\mean{\vec{E}}_{Ine}}$ and $P^{1D}_{\Sigma \boldsymbol{\tau}}/P^{1D}_{\mean{\vec{E}}_{RHS}}$. Panel a) in Figure \ref{fig:1spectrum_E_taus2}(a) shows these ratios and the color code is the same as in Figure \ref{fig:1E_taus_1D_2}. All ratios except $P^{1D}_{\boldsymbol{\tau}_{Ine}}/P^{1D}_{\mean{\vec{E}}_{Ine}}$ show an approximately constant value $\sim 10^{-4}$ for $kd_{i}<0.7$ and then steadily increases up to $kd_{i} = 10$ (the electron inertia scale). This is indicative of the negligible energy in the anomalous terms at large scales and the progressive exchange of spectral energy at smaller scales. Conversely, $P^{1D}_{\boldsymbol{\tau}_{Ine}}/P^{1D}_{\mean{\vec{E}}_{Ine}}$ show a slow decrease for $kd_{i}<1$, followed by an approximately constant value between $1<kd_{i}<7$ and a slight increase $7<kd_{i}<10$. 

To explore the level of alignment between anomalous and filter terms, we compute the phase difference between the anomalous contributions and their corresponding filtered term given by

\begin{eqnarray}
    \Delta \phi = \tan^{-1}{\left|\frac{Im(\tilde{\boldsymbol{\tau}}_{\nu} \cdot \tilde{\mean{\vec{E}}}_{\nu}^{*})}{Re(\tilde{\boldsymbol{\tau}}_{\nu} \cdot \tilde{\mean{\vec{E}}}_{\nu}^{*})} \right|},
\end{eqnarray}

\noindent where $\tilde{\boldsymbol{\tau}}_{\nu}$ and $\tilde{\mean{\vec{E}}}_{\nu}$ represent the Fourier transform of the anomalous and filtered terms in the aforementioned ratios respectively and $\tilde{\mean{\vec{E}}}_{\nu}^{*}$ represents the complex conjugate of the filtered terms. Using this definition, phase differences of $90^\circ$ correspond to fields that are either uncorrelated or vectorially orthogonal to each other, while $0^\circ$ and $180^\circ$ correspond to fields that are both perfectly correlated and vectorially aligned/anti-aligned.

Figure \ref{fig:1spectrum_E_taus2}(b) depicts the phase difference $\Delta \phi$ for each pair, $0^{\circ}$ represents perfect alignment and $\Delta \phi=90^{\circ}$ represents misalignment. For the MHD and Hall terms, $\Delta \phi(\boldsymbol{\tau}_{nu_{i}\times B},\mean{\vec{E}}_{MHD})$ and $\Delta \phi(\boldsymbol{\tau}_{J\times B},\mean{\vec{E}}_{Hall})$ show moderate alignment that progressively deteriorates towards smaller-scales up to $kd_{i}<k_{c}d_{i}$.  $\Delta \phi(\boldsymbol{\tau}_{nu_{i}\times B},\mean{\vec{E}}_{MHD})$ remains approximately constant with the anomalous contribution partially aligned with the resolved contribution in the range $k_{c}d_{i}<kd_{i}<5$. At smaller scales the alignment between these fields is enhanced; however, this occurs at scales well below $k_{c}d_{i}>1$, where the filter has significantly attenuated the fluctuation energy.

$\Delta \phi(\boldsymbol{\tau}_{J\times B},\mean{\vec{E}}_{Hall})$ behaves similar to $\Delta \phi(\boldsymbol{\tau}_{nu_{i}\times B},\mean{\vec{E}}_{MHD})$ for scales larger than $\Delta_c$, with the behavior of the alignments differing at smaller scales where the fluctuation energy has been attenuated by the filter. 

$\Delta \phi(\boldsymbol{\tau}_{Ine},\mean{\vec{E}}_{Ine})$ shows more alignment between the anomalous and resolved contributions than the other contributions to the electric fields for scales larger than $\Delta_c$; however, recall that the amplitude of these terms is significantly reduced relative to the other terms.

$\Delta \phi(\Sigma \boldsymbol{\tau},\mean{\vec{E}}_{RHS})$ follows $\Delta \phi(\boldsymbol{\tau}_{nu_{i}\times B},\mean{\vec{E}}_{MHD})$ and $\Delta \phi(\boldsymbol{\tau}_{J\times B},\mean{\vec{E}}_{Hall})$ up to $k d_{i}=1.8$ and the phase shift between the total anomalous and resolved electric fields approaches $90^{\circ}$ (uncorrelated) at $k_{c}$.

At large scales the phase coherence information of $\Sigma\boldsymbol{\tau}$ and $\mean{\vec{E}}_{RHS}$ is determined by the MHD and Hall terms, whereas at scales smaller than $\Delta_{c}$ the MHD and the inertia term are more aligned with their filtered counterparts than anomalous and filtered Hall terms. Interestingly at wavenumbers larger than $k_c$, $\Delta \phi(\Sigma \boldsymbol{\tau},\mean{\vec{E}}_{RHS})$ follows $\Delta \phi(\boldsymbol{\tau}_{Ine},\mean{\vec{E}}_{Ine})$; however, again, this occurs at scales where the filter has significantly attenuated the fluctuation energy.

At large scales, if $\Delta \phi < 90^{\circ}$ the anomalous terms are partially correlated to their filtered counterpart and at these scales the anomalous term still carries information from the filtered fields. The anomalous term is mainly driven by nonlinear terms with cross-scale couplings (low–high frequency) and small scales are being strained by large-scale structures. For $\Delta \phi = 90^{\circ}$ the filtered and anomalous terms are uncorrelated. The anomalous term corresponds to a combination of a wide range of phase-incoherent modes with nonlinear interactions coming from both cross-scale coupling and (high-high frequency) products.


\begin{figure*}{}
\centering
\includegraphics[width=0.7\linewidth]{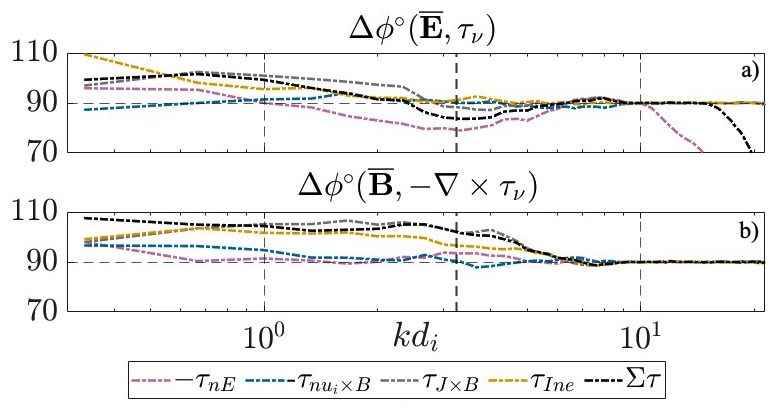}
\caption{Phase-coherence: a) Phase difference between anomalous terms and $\mean{\vec{E}}$ as a function of $k$. b) Phase difference ($\Delta \phi$) between $-\nabla \times \boldsymbol{\tau}_{\nu}$ and $\mean{\vec{B}}$.  The color convention is the same as in Figure \ref{fig:1E_taus_1D_2}. The horizontal dashed-line in both panels marks $90^{\circ}$ and the vertical dashed-lines represent the same as in Figure \ref{fig:1spectrum_E_taus2}.}
\label{fig:1spectrum_E_taus3}
\end{figure*}

In addition to establishing the level of alignment between filtered terms and their anomalous counterparts, we characterize the phase-coherence between anomalous terms and $\mean{\vec{E}}$. Figure \ref{fig:1spectrum_E_taus3}(a) shows that $\boldsymbol{\tau_{J \times B}}$ and $\boldsymbol{\tau_{Ine}}$ are moderately anti-aligned with $\mean{\vec{E}}$ progressing to no-alignment as $k$ approaches $k_c$, whilst $\Delta \phi(\mean{\vec{E}},-\boldsymbol{\tau_{nE}})$ transitions from moderated anti-alignment to alignment $k<k_{c}$. The transition occurs at $d_{i}$ and then the peak alignment occurs at $k_c$. Across scales $-\boldsymbol{\tau}_{nu_{i}\times B}$ shows no phase-coherence with $\mean{\vec{E}}$, whereas $\Sigma\boldsymbol{\tau}$ and $\mean{\vec{E}}$ are moderately anti-aligned at large scales. However, the anti-alignment is lost at $kd_{i}=2.2$ and $\Sigma\boldsymbol{\tau}$ and $\mean{\vec{E}}$ are slightly aligned around $k=k_{c}$.

In light of $\partial \mean{\vec{B}}/\partial t = - \nabla \times \mean{\vec{E}}$, only the solenoidal part of the anomalous terms contribute positively or negatively to the change in the magnetic field. Likewise, to establish whether the anomalous terms reinforce the filtered magnetic field or whether they act to oppose it (diamagnetic-like), we look at the alignment between $\mean{\vec{B}}$ and $-\nabla \times \boldsymbol{\tau}_{\nu}$. 

Figure \ref{fig:1spectrum_E_taus3}(b) shows that the phase coherence between  $- \nabla \times \Sigma\boldsymbol{\tau}$, $- \nabla \times \boldsymbol{\tau}_{J\times B}$, $- \nabla \times \boldsymbol{\tau}_{Ine}$ and $\mean{\vec{B}}$ are negative, whilst $-\nabla \times \boldsymbol{\tau_{nE}}$ and $- \nabla \times \boldsymbol{\tau_{nu_{i}\times B}}$ show almost no correlation with $\mean{\vec{B}}$.

There is a tendency for the net anomalous dynamics to diminish the resolved magnetic field on average across all scales from the largest scales to the filter scale. The net anomalous contribution is not perfectly anti-aligned, so alongside the depletion of the field, there is also a significant amount of either localized regions where the anomalous terms are either generating field or the magnetic field is also being deflected (twisted) by the anomalous electric fields. The anomalous contributions from Hall and inertial dynamics are more prone to deplete the magnetic field than the MHD and compressive dynamics. Despite displaying similar anti-correlation with the magnetic field, the relative importance of $- \nabla \times \Sigma\boldsymbol{\tau}$, $- \nabla \times \boldsymbol{\tau}_{J\times B}$, $- \nabla \times \boldsymbol{\tau}_{Ine}$ is modulated by the power in each anomalous term. Thus, $- \nabla \times \boldsymbol{\tau}_{Ine}$ has a relatively minor impact on the total phase.

The compressive dynamics accounted by $-\nabla \times \boldsymbol{\tau_{nE}}$ across almost all scales does not lead to a net change in the magnetic field or net twisting of the field. However, there could still be equal amounts positive and negative changes at a given scale that are not captured by the average over the simulation domain. At scales smaller than $d_i$, $-\nabla \times \boldsymbol{\tau_{nu_{i}\times B}}$, is also either not leading to a net change in the magnetic field  at a given scale or acting to twist the field, whereas at MHD scales it tends to lead to the depletion of the field.

\subsection{Relative Importance of the Filtered and Anomalous Terms}
\label{subsec:relative_importance}


Up to this point we have explored the relative contribution from anomalous terms to the electric field and the generation of magnetic field for the $\beta_{i}=0.25$ case, using $\Delta_{c} = 0.96d_{i}$ and at the time of maximum dissipation. We now explore how the relative contribution from anomalous terms to the electric field averaged over the turbulent domain depend on time and $\Delta_{c}$. 

\begin{figure*}{}
\centering
 \includegraphics[width=1.0\linewidth]{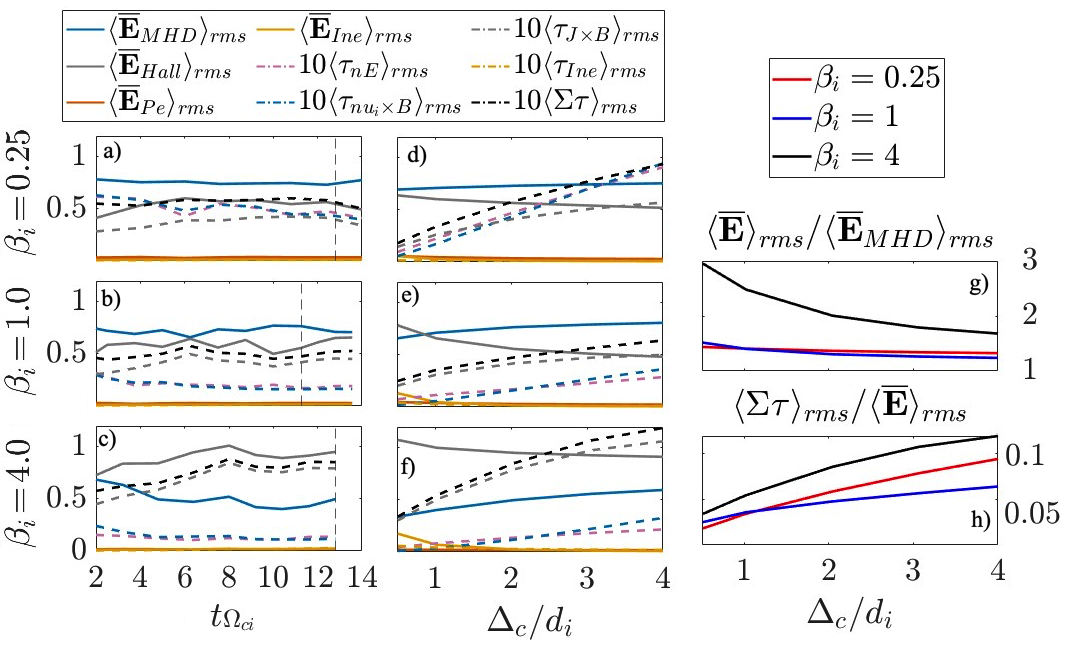}
\caption{Root-mean-square of the filtered, $\langle \mean{\vec{E}}_{MHD} \rangle_{rms}$, $\langle \mean{\vec{E}}_{Hall} \rangle_{rms}$, $\langle \mean{\vec{E}}_{P_{e}} \rangle_{rms}$, $\langle \mean{\vec{E}}_{Ine}\rangle_{rms}$, and anomalous, $\langle \boldsymbol{\tau}_{nE} \rangle_{rms}$, $\langle \boldsymbol{\tau}_{nu_{i} \times B} \rangle_{rms}$, $\langle \boldsymbol{\tau}_{J \times B} \rangle_{rms}$, $\langle \boldsymbol{\tau}_{Ine} \rangle_{rms}$, $\langle \Sigma \boldsymbol{\tau} \rangle_{rms}$ contributions normalized to $\langle \mean{\vec{E}} \rangle_{rms}$. The anomalous terms are scaled by 10 for visualization purposes. Panels a), b) and c) show the dependence as a function of time for $\beta_{i}=0.25,1,4$ respectively. The vertical dashed-line marks the time of maximum turbulent activity. Panels d), e) and f) show the filter-scale dependence at the time of maximum turbulent activity for $\beta_{i}=0.25,1,4$ respectively. Panel g) shows the ratio $\langle \mean{\vec{E}} \rangle_{rms}/\langle \mean{\vec{E}}_{MHD} \rangle_{rms}$ for $\beta_{i}=0.25$ (red), $\beta_{i}=1$ (blue) and $\beta_{i}=4$ (black). Panel g) shows the ratio $\langle \Sigma \boldsymbol{\tau} \rangle_{rms}/\langle \mean{\vec{E}} \rangle_{rms}$ for $\beta_{i}=0.25$ (red), $\beta_{i}=1$ (blue) and $\beta_{i}=4$ (black).}
\label{fig:composition_betas_windows}
\end{figure*}

Figure \ref{fig:composition_betas_windows}(a-c) show the root-mean-square (rms) $\langle \vec{F} \rangle_{rms} = \sqrt{ \langle F_{x}^{2} + F_{y}^{2} + F_{z}^{2} \rangle}$, of the filtered (solid-lines) and anomalous (dot-dashed-lines) terms on the generalized Ohm's law Eq. (\ref{eqn:ohm_filterd_3}), normalized to $\langle \mean{\vec{E}} \rangle_{rms}$ as a function of time (in units of the inverse of the ion gyro-frequency $\Omega_{ci} = eB/m_{i}$) for cases $\beta_{i}=0.25$, $\beta_{i}=1$ and $\beta_{i}=4$, respectively. For this analysis we set the filter cut-off to $\Delta_{c} = 2.04 d_{i}$ for the filer-scale to separate between scales above and below MHD scales. The anomalous terms are scaled by a factor of 10 for visualization purposes and the color convention is the same as in Figure \ref{fig:1E_taus_1D_2}. The time of maximum dissipation, estimated as the peak of the rms of the electric current density, is marked by a vertical dashed-line in Figure \ref{fig:composition_betas_windows}(a-c) for each simulation.

For all $\beta_{i}$ cases, the contributions from $\langle \mean{\vec{E}}_{P_{e}} \rangle_{rms}$ and $  \langle \mean{\vec{E}}_{Ine} \rangle_{rms}$ are, at least, one order of magnitude lower than $\langle \mean{\vec{E}}_{MHD} \rangle_{rms}$ at all times. For $\beta_{i} = 0.25$, $\langle \mean{\vec{E}}_{MHD} \rangle_{rms}$ and $\langle \mean{\vec{E}}_{Hall} \rangle_{rms}$ are the main contributions to the total electric field with $\langle \mean{\vec{E}}_{Hall} \rangle_{rms}< \langle \mean{\vec{E}}_{MHD} \rangle_{rms}$ at all times. 

The anomalous contributions from $\langle \boldsymbol{\tau}_{nE} \rangle_{rms}$ and $\langle \boldsymbol{\tau}_{nu_{i}\times B} \rangle_{rms}$ overlap for all times. At $t\Omega_{ci}=6$ and from $t\Omega_{ci}=11$ to $t\Omega_{ci}=14$, $\langle \boldsymbol{\tau}_{J\times B} \rangle_{rms}$ is close to $\langle \boldsymbol{\tau}_{nE} \rangle_{rms}$. 

The anomalous contribution from $\langle \boldsymbol{\tau}_{Ine} \rangle_{rms}$ is negligible and $\langle \Sigma \boldsymbol{\tau} \rangle_{rms}$ remains approximately constant contributing to $5 \%$ of the total electric field. However, since $\langle \Sigma \boldsymbol{\tau} \rangle_{rms}$ is a box-averaged quantity, it hide the local importance of $\Sigma \boldsymbol{\tau}$ in certain regions within the simulation domain.
 
For $\beta_{i} = 1$, Figure \ref{fig:composition_betas_windows}(b), although $\langle \mean{\vec{E}}_{MHD} \rangle_{rms} > \langle \mean{\vec{E}}_{Hall} \rangle_{rms}$ at all times, they show anti-correlation between their relative contributions. Unlike the $\beta_{i}=0.25$ case, the anomalous contribution is dominated by the $\langle \boldsymbol{\tau}_{J\times B} \rangle_{rms}$ contributing to $\sim 4\%$, whereas $\langle \boldsymbol{\tau}_{nu_{i}\times B} \rangle_{rms}$ and $\langle \boldsymbol{\tau}_{nE} \rangle_{rms}$ contribute to $ \leq 2\%$ of the total electric field.    

For $\beta_{i} = 4$, Figure \ref{fig:composition_betas_windows}(c), unlike previous cases, $\langle \mean{\vec{E}}_{Hall} \rangle_{rms}$ dominates the filtered contribution to the electric field as this term is progressively larger than $\langle \mean{\vec{E}}_{MHD} \rangle_{rms}$. Likewise, the anomalous contribution is dominated by $\langle \boldsymbol{\tau}_{J\times B} \rangle_{rms}$ and both reach a plateau $\sim 8\%$ contribution from $t\Omega_{ci} = 8$. Note that the time of saturation of $\langle \Sigma \boldsymbol{\tau} \rangle_{rms}$ is $4 \Omega_{ci}^{-1}$ earlier than the time of maximum dissipation.  

We now explore the dependence of the filtered and anomalous terms with $\Delta_{c}$ in the range $0.48d_{i} \leq \Delta_{c} \leq 4.12 d_{i} $, Figure \ref{fig:composition_betas_windows}(d-f), at the time of maximum dissipation.

For all $\beta_{i}$ cases, $\langle \mean{\vec{E}}_{MHD} \rangle_{rms}$ increases with $\Delta_{c}$ whilst $\langle \mean{\vec{E}}_{Hall} \rangle_{rms}$ decreases with $\Delta_{c}$. For $\beta_{i}=0.25$ the increase in $\langle \mean{\vec{E}}_{MHD} \rangle_{rms}$ is the smallest of the three cases. Conversely, for all $\Delta_{c}$, $\langle \mean{\vec{E}}_{P_{e}} \rangle_{rms}>\langle \mean{\vec{E}}_{Ine} \rangle_{rms}$ (not shown here) and their contributions are negligible.

For $\beta_{i}=4$, $\langle \mean{\vec{E}}_{Hall} \rangle_{rms}>\langle \mean{\vec{E}}_{HMD} \rangle_{rms}$ for all $\Delta_{c}$. This is because the injection scale for this run is at scales below the ion gyroradious ($\rho_{i}=2d_{i}$ for $\beta_{i}=4$) implying that the system is not in the MHD range.  

The decrease in $\langle \mean{\vec{E}}_{Hall} \rangle_{rms}$ is expected since the energy in the Hall term is distributed across $1<kd_{i}<10$. Likewise, the increase in $\langle \mean{\vec{E}}_{MHD} \rangle_{rms}$ is consistent with most of the energy being stored in the MHD term at $kd_{i}<1$ and the small-scales contribute negatively to the MHD term. Thus, increasing the filter scale removes cancellations between the correlated fields $\vec{u}$ and $\vec{B}$.

For $\beta_{i}=0.25$, Figure \ref{fig:composition_betas_windows}(d), the anomalous terms $\langle \boldsymbol{\tau}_{nE} \rangle_{rms}$, $\langle \boldsymbol{\tau}_{nu_{i} \times B} \rangle_{rms}$ and $\langle \Sigma \boldsymbol{\tau} \rangle_{rms}$ grow linearly with $\Delta_{c}$ reaching an $8\%$ contribution to the filtered electric field at $\Delta_{c}=4d_{i}$, whereas $\langle \boldsymbol{\tau}_{J \times B} \rangle_{rms}$ starts saturating from $\Delta_{c}=2d_{i}$ towards a $5\%$ contribution. For $\beta_{i}=1$, Figure \ref{fig:composition_betas_windows}(e), $\langle \boldsymbol{\tau}_{nE} \rangle_{rms}$ and $\langle \boldsymbol{\tau}_{nu_{i} \times B} \rangle$ increase shallower with $\Delta_{c}$ compared to the previous case, whilst $\langle \Sigma \boldsymbol{\tau} \rangle_{rms}$ follows $\langle \boldsymbol{\tau}_{J \times B} \rangle_{rms}$ which increases towards $5\%$ contribution. For $\beta_{i}=4$, Figure \ref{fig:composition_betas_windows}(f), $\langle \Sigma \boldsymbol{\tau} \rangle_{rms}$ is dominated by $\langle \boldsymbol{\tau}_{J \times B} \rangle_{rms}$ and both increase rapidly with $\Delta_{c}$ reaching a $10\%$ contribution. 

At larger filter scales, the amplitude of the anomalous term $\langle \Sigma \boldsymbol{\tau} \rangle_{rms}$ matches the amplitude of the $\langle \mean{\vec{E}}_{Hall} \rangle_{rms}$ term whilst the total filtered electric field is dominated by $\langle \mean{\vec{E}}_{MHD} \rangle_{rms}$. Although it is clear that increasing $\Delta_{c}$ leads to more energy stored in the anomalous terms, it is not clear how this affects the contribution from the anomalous term to the total filtered electric field for the different $\beta_{i}$ cases.

Figure \ref{fig:composition_betas_windows}(g) shows the ratio $\langle \mean{\vec{E}} \rangle_{rms} / \langle\mean{\vec{E}}_{MHD} \rangle_{rms}$ as a function of the filter-scale for the three cases $\beta_{i} = 0.25$ (red), $\beta_{i} = 1$ (blue) and $\beta_{i} = 4$ (black). Although, the ratio $\langle \mean{\vec{E}} \rangle_{rms} / \langle \mean{\vec{E}}_{MHD} \rangle_{rms}$ falls monotonically as $\Delta_{c}$ increases for the three $\beta_{i}$, for $\beta_{i}=4$, the fall is more dramatic since $\langle \mean{\vec{E}} \rangle_{rms} / \langle \mean{\vec{E}}_{MHD} \rangle_{rms}$ is larger because the main contribution to the electric field comes from the Hall term instead of the MHD term.

For the total contribution from the anomalous terms, Figure \ref{fig:composition_betas_windows}(h) shows that the ratio $\langle \Sigma \boldsymbol{\tau} \rangle_{rms} / \langle \mean{\vec{E}} \rangle_{rms}$ increases with $\Delta_{c}$ for all $\beta_{i}$ cases as more turbulent energy is removed from the filtered fields and added to the anomalous terms. This effect is possibly enhanced for high $\beta_{i}$ by nonlinear contributions from the Hall term but not due to compressive since the term that contains the compressive effects that can potentially contribute to the anomalous electric field, $\langle \boldsymbol{\tau}_{nE} \rangle_{rms} / \langle \mean{\vec{E}} \rangle_{rms}$, is smaller at high ion plasma beta which is consistent with the incompressible high plasma beta limit associated with much faster sound speed compared to the Alfvén speed. However, for the cases with $\beta_{i} = 0.25$ and $\beta_{i} = 4$, $\langle \Sigma \boldsymbol{\tau} \rangle_{rms} / \langle \mean{\vec{E}} \rangle_{rms}$ scale roughly similar, while the case with $\beta_{i} = 1$ behaves differently. There is seemingly a suppression of the anomalous electric field for $\beta_{i}=1$.

Since the ratio $\langle \Sigma \boldsymbol{\tau} \rangle_{rms} / \langle \mean{\vec{E}} \rangle_{rms}$ does not scale linearly with $\Delta_{c}$ for any $\beta_{i}$, and considering that $\Delta \phi(\mean{\vec{E}}, \Sigma \boldsymbol{\tau})$ are partially anti-aligned despite $\Delta \phi(\mean{\vec{E}}_{RHS}, \Sigma \boldsymbol{\tau})$ being partially aligned, the nonlinear correlations between fluctuating fields and how they contribute to the total anomalous electric field is a key aspect for modeling anomalous electric fields.


\subsection{Modeling the Anomalous Electric Field}

In this section, we explore the dependence of $\Sigma\boldsymbol{\tau}$ on the filtered fields in order to provide insights into SGS models that can be applied to LES in collisionless plasmas. Our analysis is motivated by well established effects associated with the anomalous electromotive force in MHD systems \cite{brandenburg2005astrophysical,yokoi2011modeling,yokoi2016new,yokoi2023unappreciated}, with the aim of assessing the extent to which these effects can be used to describe a fully collisionless system. In these models, $\Sigma\boldsymbol{\tau}$ is a linear function of $\mean{\vec{B}}$, $\mean{\vec{J}}$, and $\mean{\boldsymbol{\Omega}}=\nabla \times \mean{\vec{u}_{i}}$ with transport coefficients parameterizing the effect of the residual turbulent fluctuations. The dependence with $\mean{\vec{B}},\mean{\vec{J}}, \mean{\boldsymbol{\Omega}}$ is motivated by mean-field dynamo theory, where the transport coefficients $C_{\alpha}, C_{\beta}$ and $C_{\gamma}$ depend on the adopted closure theory \cite{brandenburg2005astrophysical,yokoi2011modeling,yokoi2016new,yokoi2023unappreciated}. The functional form of the different terms in these models associates them with different physical effects. The term proportional to $\mean{\vec{B}}$ is often referred to as the "alpha" effect and is associated with the generation of large-scale helical fields via the small-scale turbulent dynamics \cite{moffatt1978magnetic,brandenburg2005astrophysical}. The term proportional to $\mean{\vec{J}}$ is associated with turbulent resistivity in which the small-scale turbulent motions lead to an effective diffusion of large-scale magnetic gradients \cite{krause2016mean}. Finally, the term proportional to $\mean{\boldsymbol{\Omega}}$ is associated with the the so-called cross-helicity effect in which alignments between the small-scale turbulent magnetic field and velocity enable the coupling to large-scale velocity shears or vortical motions \cite{yokoi2011modeling,yokoi2016new,yokoi2023unappreciated}.

We focus on two variants of these models, in the first model, given by

\begin{eqnarray}
 \Sigma \tau_{M1,\eta} = C_{\alpha_{1},\eta} \mean{{B}}_{\eta} + C_{\beta_{1}\eta} \mean{{J}}_{\eta} + C_{\gamma_{1},\eta} \mean{{\Omega}}_{\eta},
 \label{eqn:M1}
\end{eqnarray}

\noindent where the coefficients ${C}_{\alpha_{1},nu}$, ${C}_{\beta_{1},\eta}$, and ${C}_{\gamma_{1},\eta}$ are taken to be spatially constant throughout the domain, but potentially allowed to be anisotropic vectors and the subscript $\eta$ represent the component ($\eta = x,y,z$). In the second model, given by 

\begin{eqnarray}
\Sigma {\tau}_{M2,\eta} = C_{\alpha_{2},\eta} H \mean{{B}}_{\eta} + C_{\beta_{2},\eta} (K-K_{R})\mean{{J}}_{\eta} + C_{\gamma_{2},\eta} W \mean{{\Omega}}_{\eta},
\label{eqn:M2}
\end{eqnarray} 

\noindent the transport coefficients are allowed to vary spatially based on the fluctuation amplitudes and alignments of the residual fluctuations, such that ${C}_{\alpha_{2},\eta}$, ${ C}_{\beta_{2},\eta}$, and ${C}_{\gamma_{2},\eta}$ are constants and 

\begin{eqnarray}
    H = \mean{-\vec{u}_{i}' \cdot \boldsymbol{\omega}' + \vec{b}' \cdot \vec{j}'},\\
    K = \mean{\frac{\vec{u}_{i}'^{2} + \vec{b}'^{2}}{2}}, \\
    K_{R} = \mean{\frac{\vec{u}_{i}'^{2} - \vec{b}'^{2}}{2}} \  \text{and} \\
    W = \mean{\vec{u}_{i}' \cdot \vec{b}' }, 
    \label{eqn:rugged_things}
\end{eqnarray}

\noindent are the anomalous residual helicity, anomalous energy density, anomalous residual energy density and anomalous cross-helicity. In these definitions $\vec{u}_{i}'$ is the residual ion velocity, $\boldsymbol{\omega}' = \nabla \times \vec{u}_{i}'$ is the residual ion vorticity, $\vec{b}' = \vec{B}'/\sqrt{\mu_{0}m_{i}n_{0}}$ is the residual magnetic field in Alfvén units and $\vec{j}' = \vec{J}'\mu_{0}/\sqrt{\mu_{0}m_{i}n_{0}}$ is the residual electric current density in Alfvén units.

To build the models, we compute $\mean{\vec{J}}, \ \mean{\vec{B}}, \ \mean{\boldsymbol{\Omega}}, H, \ K, K_{R}$ and $W$ on the simulation domain for $\beta_{i}=0.25$ and $\Delta_{c}=2.04 d_{i}$. This cut-off scale ensures that the scales to be model are larger than both $d_{i}$ and $\rho_{i}$. All three $\beta$ cases produce similar qualitative results, so we focus on showing the results for the $\beta_{i}=0.25$ case. We extract the values for a random sample of one quarter of the points in the simulation domain for the three $\beta_{i}$ cases. For each $\beta_{i}$ case we use the sample and a multilinear fit of each component of $\Sigma{\boldsymbol{\tau}}$ to estimate the model coefficients for $\Sigma{\boldsymbol{\tau}}_{M1}$ and $\Sigma{\boldsymbol{\tau}}_{M2}$. The estimated values for the $\beta_{i}=0.25$ case are summarized in Table \ref{tab:models_coef}. To estimate uncertainties on the estimated coefficients, we extract 10000 independent random samples of one quarter of the points in the domain and determine the half width of the spread in coefficients obtained from each of those random samples. This  provides an estimate of how robust the extracted coefficients are. In addition, we also compute the averaged Pearson coefficient for each model, which provides an estimate of how much of the actual $\Sigma{\boldsymbol{\tau}}$ signal is captured by the model.

The coefficients $C_{\alpha_{1},\eta}$ are positive while $C_{\alpha_{2},\eta}$ can be positive or negative. Since the ``alpha'' term is associated with the twisting or unwinding of the magnetic field \cite{yokoi2016new} there is no restriction on the sign of its coefficients. The coefficients $C_{\alpha_{1},\eta}$ show strong anisotropy with respect to $\mean{\vec{B}}_{0}$ while all $C_{\alpha_{2},\eta}$ have the same order. The coefficients $C_{\beta_{1,2},\eta}$ are positive, which is consistent with a magnetic diffusivity effect ($\partial \mean{\vec{B}}/\partial t-\nabla \times \Sigma \boldsymbol{\tau} \sim C_{\beta} \nabla^{2} \mean{\vec{J}} $ ). The coefficients $C_{\beta_{1,2},z} < C_{\beta_{1,2},x,y}$ showing a strong anisotropy with respect to $\mean{\vec{B}}_{0}$. This suggests a stronger diffusive effect in the direction perpendicular to $\mean{\vec{B}}_{0}$. The coefficients $C_{\gamma_{1},\eta}$ are also positive while $C_{\gamma_{2},\eta}$ can be positive or negative. Since the ``Omega'' term is associated with change in the magnetic field due to the mean vorticity \cite{yokoi2016new} there is no restriction on the sign of its coefficients. The coefficients $C_{\gamma_{1},\eta}$ show weak anisotropy while all $C_{\gamma_{2},\eta}$ show strong anisotropy. The coefficient $C_{\gamma_{2},z}\approx 0$, is smaller than the estimated uncertainty suggesting that the ``Omega effect is negligible along $\mean{\vec{B}}$.   

These results show that using a single coefficient for the three components hides the anisotropy with respect to the magnetic field. Conversely, the Pearson coefficient for any component is lower than $0.28$ for $\Sigma \boldsymbol{\tau}_{M1}$ and lower than $0.33$ for $\Sigma \boldsymbol{\tau}_{M2}$ indicating that there is a significant component of the anomalous electric field in a fully collisionless plasma, $\Sigma \boldsymbol{\tau}$, that is not well reproduced by these simple MHD-motivated models.

\begin{table}[h!]
\centering
\renewcommand{\arraystretch}{1.2}
\begin{tabular}{c|ccc}
\hline
 & $\Sigma \boldsymbol{\tau}_{M1,x}$ & $\Sigma \boldsymbol{\tau}_{M1,y}$ & $\Sigma \boldsymbol{\tau}_{M1,z}$ \\ \hline


$C_{\alpha_{1},\eta}$ 
& $1.188 \pm 0.017$
& $0.0295 \pm 0.017$
& $0.324 \pm 0.014$ \\

$C_{\beta_{1},\eta}$  
& $2.620 \pm 0.018$ 
& $3.439 \pm 0.019$ 
& $0.367 \pm 0.016$ \\

$C_{\gamma_{1},\eta}$ 
& $0.705 \pm 0.018$
& $0.442 \pm 0.017$
& $0.899 \pm 0.014$\\

Pearson Coeff. 
& 0.243	
& 0.279	
& 0.112\\
\hline
 & $\Sigma \boldsymbol{\tau}_{M2,x}$ & $\Sigma \boldsymbol{\tau}_{M2,y}$ & $\Sigma \boldsymbol{\tau}_{M2,z}$ \\ \hline

$C_{\alpha_{2},\eta}$ 
& $-0.202 \pm 0.019$
& $-0.472 \pm 0.022$
& $0.332 \pm 0.015$ \\

$C_{\beta_{2},\eta}$  
& $2.390 \pm 0.021$
& $3.834 \pm 0.026$
& $0.427 \pm 0.022$ \\

$C_{\gamma_{2},\eta}$ 
& $0.0740 \pm 0.021$
& $0.272 \pm 0.022$
& $-0.000886 \pm 0.015$ 
\\ 

Pearson Coeff. 
& 0.214	
& 0.325	
& 0.057
\\ \hline
\end{tabular}
\caption{Model coefficients with $\pm$ half confidence interval, both scaled by $10^{-3}$ and averaged Pearson coefficient for each model from Eqs. (\ref{eqn:M1}) and (\ref{eqn:M2}) for $\beta_{i} = 0.25$.}
\label{tab:models_coef}
\end{table}

\begin{figure*}{}
\centering
 \includegraphics[width=1\linewidth]{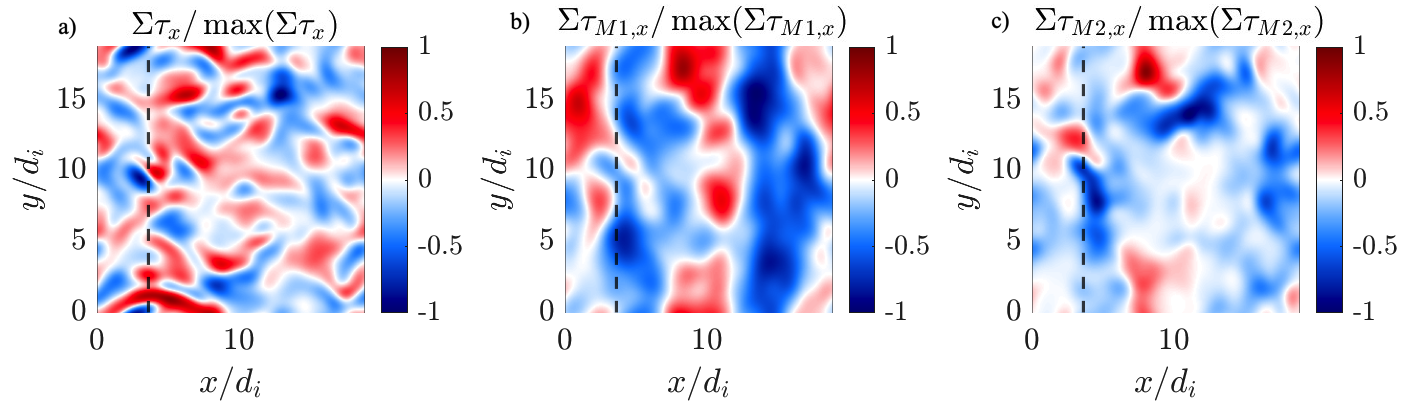}
 \caption{Comparison between anomalous term and the models. 2D cuts of x-component of the the normalized $\Sigma \tau_{x} / \max(\Sigma \tau_{x}) $ (panel a),  $\Sigma \tau_{M1,x} /  \max(\Sigma \tau_{M1,x})$ (panel b) and $\Sigma \tau_{M2,x} /  \max(\Sigma \tau_{M2,x})$ (panel c).}
\label{fig:dependence_resolved300}
\end{figure*}

Figure \ref{fig:dependence_resolved300}(a-c) depicts 2D cuts of $\Sigma \tau_{x}/max(\Sigma \tau_{x})$, $\Sigma \tau_{M1,x}/max(\Sigma \tau_{M1,x})$ and $\Sigma \tau_{M2,x}/max(\Sigma \tau_{M2,x})$, respectively, for $\beta_{i}=0.25$. Contrasting Figure \ref{fig:dependence_resolved300}(b,c) with Figure \ref{fig:dependence_resolved300}(a), both models display larger scales than the actual signal $\Sigma \boldsymbol{\tau}$. $\Sigma \boldsymbol{\tau}_{M2}$ is marginally better than $\Sigma \boldsymbol{\tau}_{M1}$ as $\Sigma \tau_{M2}$ capture better the small-scale nature of the anomalous term. However, neither of the models recover the spatial distribution of $\Sigma \boldsymbol{\tau}$. 

\begin{figure*}{}
\centering
 \includegraphics[width=0.7\linewidth]{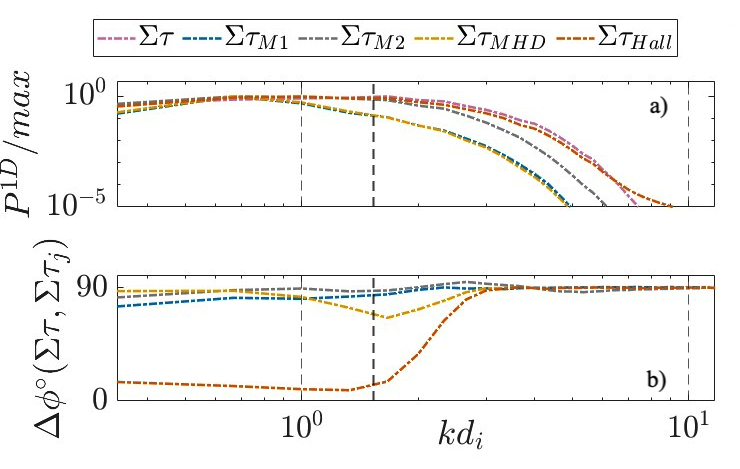}
 \caption{Spectral information. Panel a) omnidirectional power spectral density normalized to its maximum ($P^{1D}/max$) of $\Sigma \boldsymbol{\tau}$ (light purple), $\Sigma \boldsymbol{\tau}_{M1}$ (blue), $\Sigma \boldsymbol{\tau}_{M2}$ (gray), $\Sigma \boldsymbol{\tau}_{MHD}$ (yellow) and $\Sigma \boldsymbol{\tau}_{Hall}$ (orange). Panel b) phase difference $\Delta \phi$ between $\Sigma \boldsymbol{\tau}_{j}$ and $\Sigma\boldsymbol{\tau}$, for $j = M1, M2, MHD, Hall$. The thick vertical dashed-line represents the filter cutoff-scale ($k_{c}$) and the thin dashed-line marks $kd_{i}=1$.}
\label{fig:dependence_resolved3_025}
\end{figure*}

Additionally, we compute the spectrum of $\Sigma \boldsymbol{\tau}$, the models $\Sigma\boldsymbol{\tau}_{M1}$ and $\Sigma\boldsymbol{\tau}_{M2}$, as well as the spectrum of the effective anomalous terms $\Sigma \boldsymbol{\tau}_{MHD}$ Eq. (\ref{eqn:tau_MHD}) and $\Sigma \boldsymbol{\tau}_{Hall}$ Eq. (\ref{eqn:tau_Hall}). Figure \ref{fig:dependence_resolved3_025}(a) depicts the omnidirectional power spectral density of $\Sigma \boldsymbol{\tau}$, $\Sigma \boldsymbol{\tau}_{M1}$, $\Sigma \boldsymbol{\tau}_{M2}$, $\Sigma \boldsymbol{\tau}_{MHD}$ and $\Sigma \boldsymbol{\tau}_{Hall}$. We normalized to the maximum of each $P^{1D}$ to capture the shape of the spectrum.

Comparing the spectrum of $\Sigma \boldsymbol{\tau}$ with $\Sigma \boldsymbol{\tau}_{M1}$ shows that the model $\Sigma \boldsymbol{\tau}_{M1}$ underestimates the power in the anomalous electric field at smaller scales (even those which are larger than $k_c$). Conversely, the spectrum of $\Sigma\boldsymbol{\tau}_{M2}$ shows better agreement with $\Sigma \boldsymbol{\tau}$ in the shape of the spectrum up to $k_c$. This suggests that allowing the transport coefficients to vary spatially in response to the unresolved $H$, $(K-K_R)$, and $W$ retains some key information about the anomalous electric field.

The normalized spectrum of $\Sigma \boldsymbol{\tau}_{MHD}$ matches to a high degree the normalized spectrum of $\Sigma \boldsymbol{\tau}_{M1}$. On the other hand, the spectrum of $\Sigma \boldsymbol{\tau}_{Hall}$ follows the spectrum of $\Sigma \boldsymbol{\tau}$ showing that at scales larger than $\Delta_{c}$ the contribution from the filtered pressure and inertia terms are negligible, consistent with the pressure and electron inertia terms associated with the filtered fields being a negligible contribution to the spectrum as seen in Figure \ref{fig:1spectrum_E_taus}. Thus, these models capture contributions from two different scales, $\Sigma \boldsymbol{\tau}_{M1}$ captures MHD scales whereas $\Sigma \boldsymbol{\tau}_{M1}$ captures contributions at witch phenomena beyond MHD are important.

The fact that the shape of $\Sigma{\boldsymbol{\tau}}_{M1}$ and $\Sigma{\boldsymbol{\tau}}_{M2}$ spectra agree reasonably well with the $\Sigma{\boldsymbol{\tau}}_{MHD}$ and $\Sigma{\boldsymbol{\tau}}_{Hall}$ spectra, respectively up to $k_c$, suggests the two models may be capturing some element of the dynamics relevant to different scales in the flow. $\Sigma{\boldsymbol {\tau}}_{M1}$ captures, at least some element of, the cumulative effect of all terms that go beyond MHD, whereas $\Sigma{\boldsymbol{\tau}}_{M2}$ appears to be better at reproducing the shape of the spectrum for $\Sigma{\boldsymbol{\tau}}$ (which behaves similar to $\Sigma{\boldsymbol{\tau}}_{Hall}$)  up to $k_c$.

With regard to the phase coherence between the models and the anomalous contribution, Figure \ref{fig:dependence_resolved3_025}(b) shows the phase difference $\Delta \phi$ between $\Sigma \boldsymbol{\tau}_{M1}$, $\Sigma \boldsymbol{\tau}_{M2} $, $ \Sigma \boldsymbol{\tau}_{MHD}$  and $\Sigma \boldsymbol{\tau}_{Hall}$ with respect to $\Sigma \boldsymbol{\tau}$. $\Sigma \boldsymbol{\tau}_{M1}$ and $\Sigma \boldsymbol{\tau}_{M2}$ are are not well correlated with $\Sigma{\boldsymbol{\tau}}$ with $\Delta\phi \approx 90^\circ$. Similarly, the phase difference for $\Sigma \boldsymbol{\tau}_{MHD}$ shows that it is not well correlated at scales larger than $k_{c}$ and $\Delta \phi$ has a minimum close to $k_c$ but it does not become drastically more aligned with $\Sigma{\boldsymbol{\tau}}$. Conversely, $\Delta \phi$ for $\Sigma \boldsymbol{\tau}_{Hall}$ remains below $30^{\circ}$ up to $k_{c}$ after which it increases towards $\approx 90^{\circ}$. Therefore, at large scales $\Sigma \boldsymbol{\tau}$ and $\Sigma \boldsymbol{\tau}_{Hall}$ are positively correlated but the coherence at scales below $k_{c}d_{i}>3$ (below the filter scale) is lost. This implies that including the filtered pressure and inertial terms into the effective anomalous contribution does not destroy the coherence with $\Sigma \boldsymbol{\tau}$. In this sense, including the filtered Hall term into the effective anomalous contribution not only lead to an underestimation of the shape of the spectrum but it also removes the spatial coherence.   

These results suggests that the multi-linear approach motivated by MHD dynamics does not captures the full dynamics of the anomalous electric field in collisionless plasmas. Moreover, the coefficients ${\bf C}_\alpha$, ${\bf C}_\beta$, and ${\bf C}_\gamma$ have different values for different components and are, therefore, not isotropic, which is likely a manifestation of the anisotropy of the turbulence in the system.


\section{Discussion and Conclusions}
\label{sec:discussion}

In the filtered generalized Ohm's law Eq. (\ref{eqn:ohm_filterd_3}) several anomalous terms ($\boldsymbol{\tau}_{nE}, \boldsymbol{\tau}_{nu_{i}\times B}, \boldsymbol{\tau}_{J\times B}$, $\boldsymbol{\tau}_{Ine}$) can potentially contribute to the large scale electric field $\mean{\vec{E}}$.

The anomalous terms associated with the electric and density fluctuations $\boldsymbol{\tau}_{nE}$ tends to be partially balanced by $\boldsymbol{\tau}_{nu_{i} \times B}$ at the largest scales. This behavior is to be expected, as the combination of these two terms removes the compressive nonlinearity associated with the MHD electric field, resulting in $\boldsymbol{\tau}_{u_i\times B} = \mean{\vec{u}_i \times \vec{B}} - \mean{\vec{u}_i} \times \mean{\vec{B}}$.

The balance between $\boldsymbol{\tau}_{nE}$ and $\boldsymbol{\tau}_{nu_{i}\times B}$ at large scales and $\boldsymbol{\tau}_{nE}$ and $\boldsymbol{\tau}_{nu_{e} \times B}$ at small scales is a consequence of scale locality, i.e., negligible interaction between large-scale and small-scale processes, and the absence of anomalous contribution $\boldsymbol{\tau}_{u_{i}\times B}$ and $\boldsymbol{\tau}_{u_{e}\times B}$ at each scale.

At smaller scales, the anomalous Hall term $\boldsymbol{\tau}_{J \times B}$ contributes the most to the total anomalous electric field $\Sigma\boldsymbol{\tau}$. At this scales $\vec{E} = -\vec{u}_{i} \times \vec{B} + \vec{J} \times \vec{B}/ne = -\vec{u}_{e} \times \vec{B}$, the anomalous resistivity term $\boldsymbol{\tau}_{nE} = -\boldsymbol{\tau}_{nu_{e}\times B} + \boldsymbol{\tau}_{u_{e} \times B}$, is balanced by an anomalous transport at electron scales. This is consistent with the localized partial anti-correlation between $\boldsymbol{\tau}_{nE}$ and $\boldsymbol{\tau}_{J \times B}$ in Figure \ref{fig:1E_taus_1D_2}(b) and the switch of spectral power between $\boldsymbol{\tau}_{nu_{i} \times B}$ and $\boldsymbol{\tau}_{J \times B}$ at $kd_{i} \approx 2$ in Figure \ref{fig:1spectrum_E_taus}(d).

We find the compressive nonlinearities $\boldsymbol{\tau}_{nE}$, MHD $\boldsymbol{\tau}_{nu_{i}\times B}$, and Hall $\boldsymbol{\tau}_{J\times B}$ anomalous contribution to be the most significant. Moreover, the anomalous contributions inherit the scale dependence of their resolved counterparts. The Hall effect acts more on the smaller scales in the resolved range, $l < \rho_{i}$  $\boldsymbol{\tau}_{J\times B} >\boldsymbol{\tau}_{nE}, \boldsymbol{\tau}_{nu_{i}\times B} $, while the MHD and compressive term act more on the larger scales in the resolved range, $l > \rho_{i} $ $\boldsymbol{\tau}_{nE}, \boldsymbol{\tau}_{nu_{i}\times B} > \boldsymbol{\tau}_{J\times B}$ (see Figure \ref{fig:1spectrum_E_taus}(d)). This behavior does not dependent on the filter scale $\Delta_{c}$.  

The amplitude of the anomalous terms increase as the filter-scale ($\Delta_{c}$) increases and their total contribution measured by the $\langle \boldsymbol{\tau}_{\nu} \rangle_{rms}$ also increases with the $\Delta_{c}$. Since $\langle \boldsymbol{\tau}_{\nu} \rangle_{rms}$ is computed over the whole simulation domain, it captures the dependence at the larger scales. For this reason $\langle \boldsymbol{\tau}_{J \times B} \rangle_{rms} < \langle \boldsymbol{\tau}_{nu_{i} \times B} \rangle_{rms}$ for all $\Delta_{c}$ at $\beta_{i}=0.25$ and $\langle \boldsymbol{\tau}_{nu_{i} \times B}$ displays an almost flat behavior. Conversely, increasing the filter-scale reduces $\langle \vec{J}\times \vec{B} \rangle_{rms}$ because it removes the coherence of the small-scales, (see Figure \ref{fig:composition_betas_windows}(d)).

The simulations that we analyze in this study target an accurate representation of the kinetic range of scales in a collisionless plasma \cite{granier2024electron} and the role of the Hall term is dominant at kinetic scales. As $\beta_{i}$ increases, the contribution from the Hall term to the total electric field becomes more important. This may be because, since $\rho_{i}=\sqrt{\beta_{i}}d_{i}$, large $\beta_{i}$ implies bigger ion gyroradius and demagnetization of ions at larger scales relative to $d_{i}$ as well as dropping of the spectrum of the MHD term at larger scales which is then compensated by the Hall term. Additionally, the evolution of plasma turbulence depends on the ion plasma beta, (see Figures \ref{fig:composition_betas_windows}(a-c)). 

On the other hand, the change $\langle \boldsymbol{\tau}_{J \times B} \rangle_{rms} < \langle \boldsymbol{\tau}_{nu_{i} \times B} \rangle_{rms}$ for $\beta_{i}=0.25$ to $\langle \boldsymbol{\tau}_{J \times B} \rangle_{rms} > \langle \boldsymbol{\tau}_{nu_{i} \times B} \rangle_{rms}$ 
for $\beta_{i}=4$ (see Figure \ref{fig:composition_betas_windows}(d and f)) in is due to the fact that the MHD range for latter simulation is at $k\rho_{i} > 1$ which corresponds to $\Delta_{c} > 4 d_{i}$.

A dimensionless analysis shows that the ratio between the Hall and MHD terms $|\vec{J} \times \vec{B}/ne\vec{u}_{i}\times \vec{B}| \sim |\vec{k} \times \delta \vec{B}| / |\delta \vec{u}_{i}|$ and, although the three cases explored here, $\beta_{i}=0.25,1,4$, begin with the same level of magnetic fluctuations, the rms of the ion velocity $\langle \delta \vec{u}_{i} \rangle_{rms}$ at the peak of dissipation is lower for $\beta_{i}=4$ due to the compressibility \cite{lewis2023magnetospheric,granier2024electron}. This also leads to the relative contribution to $\Sigma \boldsymbol{\tau}$ from $\boldsymbol{\tau}_{J \times B}$ becoming more important than any other term for high beta cases (see Figure \ref{fig:composition_betas_windows}).

Our analysis of the phase coherence between the anomalous electric field $\Sigma\boldsymbol{\tau}$ and the electric field reconstructed from the filtered variables $\mean{\vec{E}}_{RHS}$ (see Figure \ref{fig:1spectrum_E_taus2}(b)) and between $\Sigma\boldsymbol{\tau}$ and the filtered electric field $\mean{\vec{E}}$ (see Figure \ref{fig:1spectrum_E_taus3}(a)) reveals that $\Sigma\boldsymbol{\tau}$ remains partially phase aligned with $\mean{\vec{E}}_{RHS}$  ($\Delta \phi < 90^{\circ}$), indicating that the unresolved contribution is dynamically linked to the resolved-scale plasma evolution rather than behaving as an independent fluctuation. In contrast, the phase difference between the $\Sigma\boldsymbol{\tau}$ and $\mean{\vec{E}}$ exceeds $90^\circ$, suggesting that the anomalous contribution primarily acts to modify, rather than reinforce, the large-scale electric field. 

The anomalous terms arise as subgrid-scale closure corrections associated with unresolved nonlinear correlations and represent the feedback of eliminated scales onto the resolved dynamics \cite{eyink2006breakdown,aluie2017coarse,camporeale2018coherent,cerri2020space}. The observed phase relationships further indicate that this feedback is not merely dissipative but systematically modifies the phase of the resolved electric field, providing direct spectral evidence that subgrid-scale dynamics alter the structure of the coarse-grained generalized Ohm's law. Rather than representing an additional coherent driving electric field, the anomalous contribution acts as a self-consistent turbulent correction to the generalized Ohm's law, reflecting the influence of unresolved fluctuations on the coarse-grained dynamics.

The cross-spectra analysis (see Figure \ref{fig:1spectrum_E_taus3}(b)) further reveals a partial anti-alignment between $\mean{\vec{B}}$ and the curl of the anomalous electric field contribution, $ \nabla \times \Sigma \tau$, particularly at intermediate scales where SGS effects are strongest. This phase relationship indicates that the SGS-induced electromotive contribution tends to oppose the coherent magnetic-field dynamics in the filtered induction equation, suggesting a systematic back-reaction of unresolved fluctuations on $\mean{\vec{B}}$ evolution. The fact that the anomalous electric field tends to act to deplete the magnetic field on average in the simulations is consistent with the relatively large turbulent diffusion coefficients $\bf C_{\beta_{1,2}}$. Moreover, the anomalous Hall term is the dominant term associated with the average depletion of the field, whereas the contribution from the MHD and compressive terms to the field depletion is negligible across all scales. 

Rather than acting as a passive correction, the SGS terms contribute actively to the induction process by generating an effective electromotive response that counteracts the resolved-scale magnetic structure. This behavior is consistent with an interpretation in which subgrid-scale fluctuations produce a non-ideal turbulent electromotive force that redistributes magnetic energy across scales and enhances scale-dependent magnetic field decorrelation. While this opposition resembles diamagnetic-like behavior in the sense of reducing coherent magnetic-field alignment, a more general interpretation is that SGS terms act as a turbulent closure contribution to the induction equation, mediating magnetic field reorganization through nonlinear multiscale interactions.

In the context of pure MHD, the anomalous contribution to the electric field comes solely from $\boldsymbol{\tau}_{u_{i}\times B}$ \cite{yokoi2011modeling,yokoi2023unappreciated}. Conversely, in the context of Hall MHD, the anomalous contributions come from $\boldsymbol{\tau}_{u_{i} \times B}$ and $\boldsymbol{\tau}_{J\times B}$ \cite{miura2022sub,miura2023numerical}, which combined effect corresponds to $\boldsymbol{\tau}_{u_{e}\times B}$. Therefore, a model for the effective electromotive force within the context of these reduced models only needs to consider how different mechanisms, e.g., mean-field dynamo the $\alpha$ and $\beta$ effects and turbulent cross helicity, contribute to the generation of the MHD and Hall term and not the additional collisionless electric fields. In this study we study anomalous electric fields from kinetic-scale origins. We decompose generalized Ohm's law into filtered and anomalous terms aiming to provide insights for SGS models of collisionless plasma turbulence at kinetic scales. We characterize the anomalous contribution to the electric field from all terms in the generalized Ohm's law and show that at scales $l \geq d_{i}$ the main kinetic effect required to be included comes from the Hall term specially for large $\beta_{i}$ plasmas.

We also test the accuracy of two SGS models $\boldsymbol{\tau}_{M1}$ and $\boldsymbol{\tau}_{M2}$ to convey the information of $\Sigma\boldsymbol{\tau}$ and observe that only $\boldsymbol{\tau}_{M2}$, a model that consider the effect of $K, K_{R}, H$ and $W$, is able to partially retain the spectral shape of $\Sigma\boldsymbol{\tau}$. However, the spatial coherence is not recovered suggesting that a more elaborated type of SGS modeling is required for collisionless plasmas. 

In this study we test two models, i.e., $\Sigma \boldsymbol{\tau}_{M1}$ (Eq. \ref{eqn:M1}) and $\Sigma \boldsymbol{\tau}_{M2}$ (Eq. \ref{eqn:M2}). The former represent a much simpler implementation in LES as it only depends on filtered quantities. The latter includes the anomalous quantities $H, K, K_{R}, W$ that need to be evolved dynamically \cite{yokoi2011modeling,yokoi2023unappreciated}. Model $\Sigma \boldsymbol{\tau}_{M1}$ does not retain the spectral shape of the total anomalous contribution, it only recovers the information associated with the MHD term as can be seen by comparing $\Sigma\boldsymbol{\tau}_{M1}$ with $\Sigma\boldsymbol{\tau}_{MHD}$, (see Figure \ref{fig:dependence_resolved3_025}(a)). Conversely, model $\Sigma \boldsymbol{\tau}_{M2}$, despite not fully recovering the shape of the spectrum of $\Sigma \boldsymbol{\tau}$, it shows a minor underestimation of the shape of the spectrum compared to $\Sigma \boldsymbol{\tau}_{M1}$. The phase information between the two models is similar, thus, both models present similar spatial distribution of the structures which are out of phase with the real structures in the simulation domain. Although model $\Sigma \boldsymbol{\tau}_{M2}$ \cite{yokoi2023unappreciated} is a combination of functional (dissipation through eddy viscosity) and structural (encodes turbulence geometry/topology) approaches, a caveat of this model is that it does not recover the coherence information, see Figure \ref{fig:dependence_resolved3_025}(b) encoded in $\Sigma \boldsymbol{\tau}$.   

Previous works \cite{miura2022sub} that incorporate the anomalous contribution from the Hall term, shows much better agreement with DNS than the model that only considers the MHD anomalous contribution. This is consistent with our results and including the Hall term is pivotal for more accurate modeling. 

Our results suggest that including the resolved electron pressure and resolved inertial terms does not change the large scale dynamics, however, their contributions to the anomalous terms might not be negligible since the biggest contributions to the anomalous dynamics come from $\boldsymbol{\tau}_{nE}$ and $\boldsymbol{\tau}_{nu_{i}\times B}$, and notably $\boldsymbol{\tau}_{nE}$ includes the nonlinearities associated with the electron pressure.

This work provides a framework for further examining the impact of small scale turbulent dynamics in collisionless plasmas that can be applied to further simulations of collisionless turbulence. Such simulations could include those with more complete descriptions of the electron-scale physics, which would incorporate the potential impact of additional fundamentally kinetic deformations to the electron distribution function that are not included in the hybrid Vlasov formalism, as well as kinetic simulations with larger scale separations into the MHD, which will be critical for fully constraining the extent to which the kinetic scale physics alters the fluid-scale dynamics. Additionally, similar analyses can be directly applied to multi-spacecraft measurements of actual collisionless space plasmas, such as those obtained from the Magnetospheric Multiscale (MMS) mission \cite{burch2016magnetospheric} and potentially future multi-scale missions, such as HelioSwarm \cite{klein2023helioswarm} and Plasma Observatory \cite{retino2022particle}. Notably, it has been demonstrated that nearly all of the terms in generalized Ohm's law can be directly estimated using MMS measurements in Earth's magnetosheath \cite{stawarz2021comparative,lewis2023magnetospheric} and the comparison of the anomalous electric field in those dataset to the results from this work will be the subject of future work. Such analyses can potentially supplement and provide further insights into recent observation of turbulent dynamo effect in Earth's magnetosheath \cite{voros2026turbulent}.

An alternative approach to model the SGS effects would explore contributions from second order and nonlinear terms to better model the anomalous contribution ($\sim \alpha B^{n} + \beta J^{n} + \gamma \boldsymbol{\Omega}^{n}$ ...) and nonlinear ($\alpha_{BJ} BJ + \alpha_{B\boldsymbol{\Omega}} B \boldsymbol{\Omega} + ...$) \cite{brandenburg2005astrophysical}. Further ways to improve the analysis would tackle the anisotropic nature of plasma turbulence as well as expansion effects. Moreover, note that this work tackles only the nature of anomalous electromotive force terms and their contribution to the induction equation and we do not address the modeling of the Reynolds stress tensor. 

Finally, in this work we opted for a direct approach, rather than a machine learning approach, to better understand anomalous electric fields. Nonetheless our result can be highly relevant for the increasing interest on using machine learning techniques to address the closure problem \cite{maulik2020neural,laperre2022identification,burles2025machine,miloshevich2026electron} and to incorporate effects beyond multi-fluid MHD into large scale simulations.

%
%

\section*{Open Research Section}


The datasets generated/analyzed for this study will be made available by the authors upon reasonable request, or deposited in a public repository upon publication

\section*{Conflict of Interest declaration}

The authors declare there are no conflicts of interest for this manuscript.

\acknowledgments

JAAR is supported by the Royal Society University Research Fellowship awards URF/R/251029 and URF/R1/201286. JES is supported by the Royal Society University Research Fellowship awards URF/R/251029 and URF/R1/201286. S.S.C. is supported by the French government through the National Research Agency (ANR) grant ``MiCRO” with the reference number ANR-23-CE31-001

%
%
\bibliography{agusample}

@article{eyink2006breakdown,
  title={The breakdown of Alfv{\'e}n’s theorem in ideal plasma flows: Necessary conditions and physical conjectures},
  author={Eyink, Gregory L and Aluie, Hussein},
  journal={Physica D: Nonlinear Phenomena},
  volume={223},
  number={1},
  pages={82--92},
  year={2006},
  publisher={Elsevier}
}

@article{voros2026turbulent,
  title={Turbulent dynamo in the terrestrial magnetosheath},
  author={V{\"o}r{\"o}s, Zolt{\'a}n and Roberts, Owen Wyn and Narita, Yasuhito and Yordanova, Emiliya and Nakamura, Rumi and Settino, Adriana and Schmid, Daniel and Volwerk, Martin and Simon Wedlund, Cyril L and Varsani, Ali and others},
  journal={Nature Communications},
  volume={17},
  number={1},
  pages={2909},
  year={2026},
  publisher={Nature Publishing Group UK London}
}

@article{retino2022particle,
  title={Particle energization in space plasmas: towards a multi-point, multi-scale plasma observatory},
  author={Retin{\`o}, Alessandro and Khotyaintsev, Yuri and Le Contel, Olivier and Marcucci, Maria Federica and Plaschke, Ferdinand and Vaivads, Andris and Angelopoulos, Vassilis and Blasi, Pasquale and Burch, Jim and De Keyser, Johan and others},
  journal={Experimental Astronomy},
  volume={54},
  number={2},
  pages={427--471},
  year={2022},
  publisher={Springer}
}

@article{klein2023helioswarm,
  title={HelioSwarm: a multipoint, multiscale mission to characterize turbulence},
  author={Klein, Kristopher G and Spence, Harlan and Alexandrova, Olga and Argall, Matthew and Arzamasskiy, Lev and Bookbinder, Jay and Broeren, Theodore and Caprioli, Damiano and Case, Anthony and Chandran, Benjamin and others},
  journal={Space Science Reviews},
  volume={219},
  number={8},
  pages={74},
  year={2023},
  publisher={Springer}
}

@book{krause2016mean,
  title={Mean-field magnetohydrodynamics and dynamo theory},
  author={Krause, Fritz and R{\"a}dler, K-H},
  year={2016},
  publisher={Elsevier}
}

@article{moffatt1978magnetic,
  title={Magnetic field generation in electrically conducting fluids},
  author={Moffatt, Henry Keith},
  journal={Cambridge Monographs on Mechanics and Applied Mathematics},
  year={1978}
}

@article{burles2025machine,
  title={The Machine Learning Approach to Moment Closure Relations for Plasma: A Review},
  author={Burles, Samuel and Camporeale, Enrico},
  journal={arXiv preprint arXiv:2511.22486},
  year={2025}
}

@article{laperre2022identification,
  title={Identification of high order closure terms from fully kinetic simulations using machine learning},
  author={Laperre, Brecht and Amaya, Jorge and Jamal, Sara and Lapenta, Giovanni},
  journal={Physics of Plasmas},
  volume={29},
  number={3},
  year={2022},
  publisher={AIP Publishing}
}

@article{maulik2020neural,
  title={Neural network representability of fully ionized plasma fluid model closures},
  author={Maulik, Romit and Garland, Nathan A and Burby, Joshua W and Tang, Xian-Zhu and Balaprakash, Prasanna},
  journal={Physics of Plasmas},
  volume={27},
  number={7},
  year={2020},
  publisher={AIP Publishing}
}

@article{miloshevich2026electron,
  title={Electron neural closure for turbulent magnetosheath simulations: Energy channels},
  author={Miloshevich, George and Vranckx, Luka and de Oliveira Lopes, Felipe Nathan and Dazzi, Pietro and Arr{\`o}, Giuseppe and Lapenta, Giovanni},
  journal={Physics of Plasmas},
  volume={33},
  number={1},
  year={2026},
  publisher={AIP Publishing}
}

@article{stawarz2021comparative,
  title={Comparative analysis of the various generalized Ohm's law terms in magnetosheath turbulence as observed by magnetospheric multiscale},
  author={Stawarz, Julia E and Matteini, Lorenzo and Parashar, TN and Franci, Luca and Eastwood, JP and Gonzalez, CA and Gingell, IL and Burch, JL and Ergun, RE and Ahmadi, Narges and others},
  journal={Journal of Geophysical Research: Space Physics},
  volume={126},
  number={1},
  pages={2020JA028447},
  year={2021},
  publisher={Wiley Online Library}
}

@article{zhong2025electromagnetic,
  title={Electromagnetic viscosity supported anomalous electric field in the electron diffusion region of collisionless magnetic reconnection},
  author={Zhong, ZH and Zhou, M and Graham, Daniel B and Pang, Y and Khotyaintsev, Yu V and Song, LJ and Li, HM and Tang, RX and Deng, XH},
  journal={Nature Communications},
  volume={16},
  number={1},
  pages={10519},
  year={2025},
  publisher={Nature Publishing Group UK London}
}

@article{lewis2023magnetospheric,
  title={Magnetospheric Multiscale measurements of turbulent electric fields in earth's magnetosheath: How do plasma conditions influence the balance of terms in generalized Ohm's law?},
  author={Lewis, Harry C and Stawarz, Julia E and Franci, Luca and Matteini, Lorenzo and Klein, Kristopher and Salem, Chadi S and Burch, James L and Ergun, Robert E and Giles, Barbara L and Russell, Christopher T and others},
  journal={Physics of Plasmas},
  volume={30},
  number={8},
  year={2023},
  publisher={AIP Publishing}
}

@article{stanish2025turbulent,
  title={On turbulent magnetic reconnection: fast and slow mean steady states},
  author={Stanish, Sage and MacTaggart, David},
  journal={Journal of Plasma Physics},
  volume={91},
  number={2},
  pages={E49},
  year={2025},
  publisher={Cambridge University Press}
}

@incollection{leonard1975energy,
  title={Energy cascade in large-eddy simulations of turbulent fluid flows},
  author={Leonard, Athony},
  booktitle={Advances in geophysics},
  volume={18},
  pages={237--248},
  year={1975},
  publisher={Elsevier}
}

@article{pope2001turbulent,
  title={Turbulent flows},
  author={Pope, Stephen B},
  journal={Measurement Science and Technology},
  volume={12},
  number={11},
  pages={2020--2021},
  year={2001}
}

@article{camporeale2018coherent,
  title={Coherent structures and spectral energy transfer in turbulent plasma: a space-filter approach},
  author={Camporeale, Enrico and Sorriso-Valvo, Luca and Califano, Francesco and Retin{\`o}, Alessandro},
  journal={Physical review letters},
  volume={120},
  number={12},
  pages={125101},
  year={2018},
  publisher={APS}
}

@article{cerri2020space,
  title={Space-filter techniques for quasi-neutral hybrid-kinetic models},
  author={Cerri, SS and Camporeale, E},
  journal={Physics of Plasmas},
  volume={27},
  number={8},
  year={2020},
  publisher={AIP Publishing}
}

@article{charbonneau2020dynamo,
  title={Dynamo models of the solar cycle},
  author={Charbonneau, Paul},
  journal={Living Reviews in Solar Physics},
  volume={17},
  number={1},
  pages={4},
  year={2020},
  publisher={Springer}
}

@article{charbonneau2010dynamo,
  title={Dynamo models of the solar cycle},
  author={Charbonneau, Paul},
  journal={Living Reviews in Solar Physics},
  volume={7},
  number={1},
  pages={1--91},
  year={2010},
  publisher={Springer}
}

@article{brandenburg2001inverse,
  title={The inverse cascade and nonlinear alpha-effect in simulations of isotropic helical hydromagnetic turbulence},
  author={Brandenburg, Axel},
  journal={The Astrophysical Journal},
  volume={550},
  number={2},
  pages={824--840},
  year={2001}
}

@article{brandenburg2018advances,
  title={Advances in mean-field dynamo theory and applications to astrophysical turbulence},
  author={Brandenburg, Axel},
  journal={Journal of Plasma Physics},
  volume={84},
  number={4},
  pages={735840404},
  year={2018},
  publisher={Cambridge University Press}
}

@article{guan2023learning,
  title={Learning physics-constrained subgrid-scale closures in the small-data regime for stable and accurate LES},
  author={Guan, Yifei and Subel, Adam and Chattopadhyay, Ashesh and Hassanzadeh, Pedram},
  journal={Physica D: Nonlinear Phenomena},
  volume={443},
  pages={133568},
  year={2023},
  publisher={Elsevier}
}

@book{sagaut2006large,
  title={Large eddy simulation for incompressible flows: an introduction},
  author={Sagaut, Pierre},
  year={2006},
  publisher={Springer}
}

@article{yokoi2016new,
  title={A new simple dynamo model for stellar activity cycle},
  author={Yokoi, Nobumitsu and Schmitt, Dieter and Pipin, Valery and Hamba, Fujihiro},
  journal={The Astrophysical Journal},
  volume={824},
  number={2},
  pages={67},
  year={2016},
  publisher={IOP Publishing}
}

@article{brandenburg2005astrophysical,
  title={Astrophysical magnetic fields and nonlinear dynamo theory},
  author={Brandenburg, Axel and Subramanian, Kandaswamy},
  journal={Physics Reports},
  volume={417},
  number={1-4},
  pages={1--209},
  year={2005},
  publisher={Elsevier}
}

@article{burch2016magnetospheric,
  title={Magnetospheric multiscale overview and science objectives},
  author={Burch, JL and Moore, TE and Torbert, RB and Giles, BL-https},
  journal={Space Science Reviews},
  volume={199},
  number={1},
  pages={5--21},
  year={2016},
  publisher={Springer}
}

@article{alexakis2022local,
  title={Local fluxes in magnetohydrodynamic turbulence},
  author={Alexakis, Alexandros and Chibbaro, Sergio},
  journal={Journal of Plasma Physics},
  volume={88},
  number={5},
  pages={905880515},
  year={2022},
  publisher={Cambridge University Press}
}

@article{theobald1994subgrid,
  title={A subgrid-scale resistivity for magnetohydrodynamics},
  author={Theobald, Michael L and Fox, Peter A and Sofia, Sabatino},
  journal={Physics of plasmas},
  volume={1},
  number={9},
  pages={3016--3032},
  year={1994},
  publisher={American Institute of Physics}
}

@article{schmidt2011fluid,
  title={A fluid-dynamical subgrid scale model for highly compressible astrophysical turbulence},
  author={Schmidt, Wolfram and Federrath, Christoph},
  journal={Astronomy \& Astrophysics},
  volume={528},
  pages={A106},
  year={2011},
  publisher={EDP Sciences}
}

@article{agullo2001large,
  title={Large eddy simulation of decaying magnetohydrodynamic turbulence with dynamic subgrid-modeling},
  author={Agullo, Olivier and M{\"u}ller, W-C and Knaepen, Bernard and Carati, Daniele},
  journal={Physics of Plasmas},
  volume={8},
  number={7},
  pages={3502--3505},
  year={2001},
  publisher={American Institute of Physics}
}

@article{miesch2015large,
  title={Large-eddy simulations of magnetohydrodynamic turbulence in heliophysics and astrophysics},
  author={Miesch, Mark and Matthaeus, William and Brandenburg, Axel and Petrosyan, Arakel and Pouquet, Annick and Cambon, Claude and Jenko, Frank and Uzdensky, Dmitri and Stone, James and Tobias, Steve and others},
  journal={Space Science Reviews},
  volume={194},
  number={1},
  pages={97--137},
  year={2015},
  publisher={Springer}
}

@article{chernyshov2007development,
  title={Development of large eddy simulation for modeling of decaying compressible magnetohydrodynamic turbulence},
  author={Chernyshov, AA and Karelsky, KV and Petrosyan, AS1146},
  journal={Physics of Fluids},
  volume={19},
  number={5},
  year={2007},
  publisher={AIP Publishing}
}

@article{chernyshov2006large,
  title={Large-eddy simulation of magnetohydrodynamic turbulence in compressible fluid},
  author={Chernyshov, AA and Karelsky, KV and Petrosyan, AS},
  journal={Physics of plasmas},
  volume={13},
  number={3},
  year={2006},
  publisher={AIP Publishing}
}

@article{shimomura1991large,
  title={Large eddy simulation of magnetohydrodynamic turbulent channel flows under a uniform magnetic field},
  author={Shimomura, Yutaka},
  journal={Physics of Fluids A: Fluid Dynamics},
  volume={3},
  number={12},
  pages={3098--3106},
  year={1991},
  publisher={American Institute of Physics}
}

@article{zank2017theory,
  title={Theory and transport of nearly incompressible magnetohydrodynamic turbulence},
  author={Zank, GP and Adhikari, L and Hunana, P and Shiota, D and Bruno, R and Telloni, D},
  journal={The Astrophysical Journal},
  volume={835},
  number={2},
  pages={147},
  year={2017},
  publisher={IOP Publishing}
}

@inproceedings{usmanov2009mhd,
  title={An MHD solar wind model with turbulence transport},
  author={Usmanov, Arcadi V and Matthaeus, William H and Breech, Ben and Goldstein, Melvyn L},
  booktitle={Numerical Modeling of Space Plasma Flows: ASTRONUM-2008},
  volume={406},
  pages={160},
  year={2009}
}

@article{zank1996evolution,
  title={Evolution of turbulent magnetic fluctuation power with heliospheric distance},
  author={Zank, GP and Matthaeus, WH and Smith, CW},
  journal={Journal of Geophysical Research: Space Physics},
  volume={101},
  number={A8},
  pages={17093--17107},
  year={1996},
  publisher={Wiley Online Library}
}

@article{matthaeus1994evolution,
  title={Evolution of energy-containing turbulent eddies in the solar wind},
  author={Matthaeus, William H and Oughton, Sean and Pontius Jr, Duane H and Zhou, Ye},
  journal={Journal of Geophysical Research: Space Physics},
  volume={99},
  number={A10},
  pages={19267--19287},
  year={1994},
  publisher={Wiley Online Library}
}

@article{marsch1989dynamics,
  title={Dynamics of correlation functions with Els{\"a}sser variables for inhomogeneous MHD turbulence},
  author={Marsch, Eckart and Tu, C-Y},
  journal={Journal of Plasma Physics},
  volume={41},
  number={3},
  pages={479--491},
  year={1989},
  publisher={Cambridge University Press}
}

@article{zhou1989non,
  title={Non-WKB evolution of solar wind fluctuations: A turbulence modeling approach},
  author={Zhou, Ye and Matthaeus, William H},
  journal={Geophysical Research Letters},
  volume={16},
  number={7},
  pages={755--758},
  year={1989},
  publisher={Wiley Online Library}
}

@article{zank2011transport,
  title={The transport of low-frequency turbulence in astrophysical flows. I. Governing equations},
  author={Zank, GP and Dosch, A and Hunana, P and Florinski, V and Matthaeus, WH and Webb, GM},
  journal={The Astrophysical Journal},
  volume={745},
  number={1},
  pages={35},
  year={2011},
  publisher={IOP Publishing}
}

@inproceedings{usmanov1996global,
  title={A global 3-D MHD solar wind model with Alfv{\'e}n waves},
  author={Usmanov, AV},
  booktitle={AIP Conference Proceedings},
  volume={382},
  number={1},
  pages={141--144},
  year={1996},
  organization={American Institute of Physics}
}

@article{hollweg1973alfven,
  title={Alfv{\'e}n waves in the solar wind: Wave pressure, Poynting flux, and angular momentum},
  author={Hollweg, Joseph V},
  journal={Journal of Geophysical Research},
  volume={78},
  number={19},
  pages={3643--3652},
  year={1973},
  publisher={Wiley Online Library}
}

@article{alazraki1971solar,
  title={Solar wind accejeration caused by the gradient of Alfven wave pressure},
  author={Alazraki, G and Couturier, P},
  journal={Astronomy and Astrophysics, Vol. 13, p. 380 (1971)},
  volume={13},
  pages={380},
  year={1971}
}

@article{parker1965dynamical,
  title={Dynamical theory of the solar wind},
  author={Parker, EN},
  journal={Space Science Reviews},
  volume={4},
  number={5},
  pages={666--708},
  year={1965},
  publisher={Springer}
}

@article{wang2022conservation,
  title={On the conservation of turbulence energy in turbulence transport models},
  author={Wang, B-B and Zank, Gary P and Adhikari, Laxman and Zhao, L-L},
  journal={The Astrophysical Journal},
  volume={928},
  number={2},
  pages={176},
  year={2022},
  publisher={IOP Publishing}
}

@article{belcher1971alfvenic,
  title={Alfv{\'e}nic wave pressures and the solar wind},
  author={Belcher, JW},
  journal={Astrophysical Journal, vol. 168, p. 509},
  volume={168},
  pages={509},
  year={1971}
}

@article{usmanov2025unified,
  title={A Unified Three-dimensional Magnetohydrodynamic Model of the Solar Corona, Solar Wind, and Global Heliosphere with Turbulence Transport},
  author={Usmanov, Arcadi V and Chhiber, Rohit and Matthaeus, William H and Roy, Sohom and Goldstein, Melvyn L},
  journal={The Astrophysical Journal},
  volume={993},
  number={1},
  pages={87},
  year={2025},
  publisher={IOP Publishing}
}

@article{dawson1983particle,
  title={Particle simulation of plasmas},
  author={Dawson, John M},
  journal={Reviews of modern physics},
  volume={55},
  number={2},
  pages={403},
  year={1983},
  publisher={APS}
}

@book{smagorinsky1993,
    author = {Smagorinsky, Joseph},
    title = {Large Eddy Simulation of Complex Engineering and Geophysical Flows},
    publisher = {Cambridge University Press},
    year = {1993}
}

@article{reynolds1895iv,
  title={IV. On the dynamical theory of incompressible viscous fluids and the determination of the criterion},
  author={Reynolds, Osborne},
  journal={Philosophical transactions of the royal society of london.(a.)},
  number={186},
  pages={123--164},
  year={1895},
  publisher={The Royal Society London}
}

@article{graham2022direct,
  title={Direct observations of anomalous resistivity and diffusion in collisionless plasma},
  author={Graham, DB and Khotyaintsev, Yu V and Andr{\'e}, M and Vaivads, Andris and Divin, A and Drake, JF and Norgren, C and Le Contel, O and Lindqvist, P-A and Rager, AC and others},
  journal={Nature Communications},
  volume={13},
  number={1},
  pages={2954},
  year={2022},
  publisher={Nature Publishing Group UK London}
}

@article{yokoi2023unappreciated,
  title={Unappreciated cross-helicity effects in plasma physics: anti-diffusion effects in dynamo and momentum transport},
  author={Yokoi, Nobumitsu},
  journal={Reviews of Modern Plasma Physics},
  volume={7},
  number={1},
  pages={33},
  year={2023},
  publisher={Springer}
}

@article{sarghini1999scale,
  title={Scale-similar models for large-eddy simulations},
  author={Sarghini, Fabrizio and Piomelli, U and Balaras, E},
  journal={Physics of Fluids},
  volume={11},
  number={6},
  pages={1596--1607},
  year={1999},
  publisher={American Institute of Physics}
}

@inproceedings{bardina1980improved,
  title={Improved subgrid-scale models for large-eddy simulation},
  author={Bardina, Jorge and Ferziger, J and Reynolds, WC},
  booktitle={13th fluid and plasmadynamics conference},
  pages={1357},
  year={1980}
}

@article{piomelli1999large,
  title={Large-eddy simulation: achievements and challenges},
  author={Piomelli, Ugo},
  journal={Progress in aerospace sciences},
  volume={35},
  number={4},
  pages={335--362},
  year={1999},
  publisher={Elsevier}
}

@article{zhiyin2015large,
  title={Large-eddy simulation: Past, present and the future},
  author={Zhiyin, Yang},
  journal={Chinese journal of Aeronautics},
  volume={28},
  number={1},
  pages={11--24},
  year={2015},
  publisher={Elsevier}
}

@article{mason1994large,
  title={Large-eddy simulation: A critical review of the technique},
  author={Mason, Paul J},
  journal={Quarterly Journal of the Royal Meteorological Society},
  volume={120},
  number={515},
  pages={1--26},
  year={1994},
  publisher={Wiley Online Library}
}

@article{dong2022reconnection,
  title={Reconnection-driven energy cascade in magnetohydrodynamic turbulence},
  author={Dong, Chuanfei and Wang, Liang and Huang, Yi-Min and Comisso, Luca and Sandstrom, Timothy A and Bhattacharjee, Amitava},
  journal={Science Advances},
  volume={8},
  number={49},
  pages={eabn7627},
  year={2022},
  publisher={American Association for the Advancement of Science}
}

@article{valentini2007hybrid,
  title={A hybrid-Vlasov model based on the current advance method for the simulation of collisionless magnetized plasma},
  author={Valentini, Francesco and Tr{\'a}vn{\'\i}{\v{c}}ek, P and Califano, Francesco and Hellinger, Petr and Mangeney, Andr{\'e}},
  journal={Journal of Computational Physics},
  volume={225},
  number={1},
  pages={753--770},
  year={2007},
  publisher={Elsevier}
}

@article{zhou1990transport,
  title={Transport and turbulence modeling of solar wind fluctuations},
  author={Zhou, Ye and Matthaeus, William H},
  journal={Journal of Geophysical Research: Space Physics},
  volume={95},
  number={A7},
  pages={10291--10311},
  year={1990},
  publisher={Wiley Online Library}
}

@article{chhiber2021large,
  title={Large-scale structure and turbulence transport in the inner solar wind: comparison of Parker Solar Probe’s first five orbits with a global 3D Reynolds-averaged MHD model},
  author={Chhiber, Rohit and Usmanov, Arcadi V and Matthaeus, William H and Goldstein, Melvyn L},
  journal={The Astrophysical Journal},
  volume={923},
  number={1},
  pages={89},
  year={2021},
  publisher={IOP Publishing}
}

@article{aluie2017coarse,
  title={Coarse-grained incompressible magnetohydrodynamics: analyzing the turbulent cascades},
  author={Aluie, Hussein},
  journal={New Journal of Physics},
  volume={19},
  number={2},
  pages={025008},
  year={2017},
  publisher={IOP Publishing}
}

@article{usmanov2016four,
  title={A four-fluid MHD model of the solar wind/interstellar medium interaction with turbulence transport and pickup protons as separate fluid},
  author={Usmanov, Arcadi V and Goldstein, Melvyn L and Matthaeus, William H},
  journal={The Astrophysical Journal},
  volume={820},
  number={1},
  pages={17},
  year={2016},
  publisher={IOP Publishing}
}

@article{howes2024fundamental,
  title={The fundamental parameters of astrophysical plasma turbulence and its dissipation: non-relativistic limit},
  author={Howes, Gregory G},
  journal={Journal of Plasma Physics},
  volume={90},
  number={5},
  pages={905900504},
  year={2024},
  publisher={Cambridge University Press}
}

@article{pouquet2022helical,
  title={Helical fluid and (Hall)-MHD turbulence: a brief review},
  author={Pouquet, Annick and Yokoi, Nobumitsu},
  journal={Philosophical Transactions of the Royal Society A},
  volume={380},
  number={2219},
  pages={20210087},
  year={2022},
  publisher={The Royal Society}
}

@article{yokoi2011modeling,
  title={Modeling the turbulent cross-helicity evolution: production, dissipation, and transport rates},
  author={Yokoi, Nobumitsu},
  journal={Journal of Turbulence},
  number={12},
  pages={N27},
  year={2011},
  publisher={Taylor \& Francis}
}

@article{miura2023numerical,
  title={Numerical Simulations of Hall MHD Turbulence with Magnetization},
  author={Miura, Hideaki and Hamba, Fujihiro},
  journal={Plasma and Fusion Research},
  volume={18},
  pages={2401022--2401022},
  year={2023},
  publisher={The Japan Society of Plasma Science and Nuclear Fusion Research}
}

@article{miura2022sub,
  title={Sub-grid-scale model for studying Hall effects on macroscopic aspects of magnetohydrodynamic turbulence},
  author={Miura, Hideaki and Hamba, Fujihiro},
  journal={Journal of Computational Physics},
  volume={448},
  pages={110692},
  year={2022},
  publisher={Elsevier}
}

@article{bruno2013solar,
  title={The solar wind as a turbulence laboratory},
  author={Bruno, Roberto and Carbone, Vincenzo},
  journal={Living Reviews in Solar Physics},
  volume={10},
  number={1},
  pages={2},
  year={2013},
  publisher={Springer}
}

@article{granier2024electron,
  title={Electron-only reconnection and ion heating in 3D3V hybrid-Vlasov plasma turbulence},
  author={Granier, Camille and Cerri, SS and Jenko, F},
  journal={The Astrophysical Journal},
  volume={974},
  number={1},
  pages={11},
  year={2024},
  publisher={IOP Publishing}
}

\end{document}